\def\doctype{0}
\def\flagreview{0}

\ifcase\doctype
  \documentclass{article}
  \usepackage{arxiv}
\or
  \documentclass[times,sageh]{sagej}
\or
  \documentclass[linenum]{SSA-BSSA}
\or
  \documentclass[linenum]{SSA-SRL}
\or
  \documentclass[linenum]{SSA-TSR}
\else
  \errmessage{Invalid value for \doctype}
\fi

\ifnum\doctype<2
    \usepackage[utf8]{inputenc} 
    \usepackage[T1]{fontenc}    
    \usepackage{geometry}       
    \usepackage{hyperref}       
    \usepackage{nameref}        
    \usepackage{url}            
    \usepackage{microtype}      
    \usepackage{nicefrac}       
    \usepackage{amssymb}
    \usepackage{booktabs}       
    \usepackage{amsfonts}       
    \usepackage{amsmath}
    \usepackage{natbib}         
    \usepackage{doi}
    \usepackage{cleveref}
    \usepackage{graphicx} 
    \usepackage{multirow}
    \usepackage{float}
    \usepackage{caption}
    \usepackage{subcaption}
    \usepackage{makecell}
\fi
\ifnum\doctype>1
    \usepackage{caption}
    \usepackage{subcaption}
\fi

\if\doctype0
    \usepackage{lineno}
    \usepackage{pdflscape} 
    \usepackage{amsmath}
    \usepackage{graphicx} 
    \usepackage{multirow}
    \usepackage{float}
    \usepackage{geometry}
\fi

\usepackage{forest}          

\if\doctype0 
\fi
\if\doctype1 
    \newcommand\BibTeX{{\rmfamily B\kern-.05em \textsc{i\kern-.025em b}\kern-.08em
    T\kern-.1667em\lower.7ex\hbox{E}\kern-.125emX}}
\fi

\title{Extracting and Characterizing Near-fault Effects from Strong Motion Observations of the 2022 Chishang Earthquake Sequence}
\if\doctype0
    \renewcommand{\shorttitle}{Extracting and Characterizing Near-Fault Effects from the 2022 Chishang Earthquake: Methodology and Validation} 
    \newcommand{\headeright}{}
    \newcommand{\undertitle}{Submitted for Consideration to BSSA}
    \hypersetup{
        pdftitle={2022 Chishang Earthquake},
        pdfsubject={physics.geo-ph, stat.AP},
        pdfauthor={G. Lavrentiadis, D. Asimaki, T.J. Carey, H.B. Mason},
        pdfkeywords={near-field ground motions, baseline correction, 2022 Chishang Earthquake},
    }
\fi
\if\doctype1 
    \runninghead{Lavrentiadis et al.}
\fi
\ifnum\doctype>1ƒ
  \ifnum\doctype<5
    \citethisauthor{Lavrentiadis, G., D. Asimaki, T.J. Carey, and H.B. Mason, and A.}
    \doi{00.0000/0000000000}
    \recdate{00 Month 0000}
  \fi
\fi

\if\doctype0
    \author{
        \href{https://orcid.org/0000-0001-6546-1340}{\includegraphics[scale=0.06]{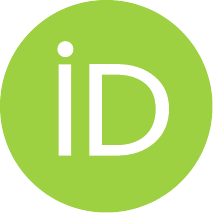}\hspace{1mm}Grigorios Lavrentiadis} \\
        Division of Engineering and Applied Sciences\\
        California Institute of Technology\\
        Pasadena, California, U.S.A. \\
        Department of Civil, Structural and Environmental Engineering \\
        State University of New York at Buffalo \\
        Buffalo, New York, U.S.A. \\
    	\texttt{glavrent@caltech.edu, glavrent@buffalo.edu} \\
        \And
        \href{https://orcid.org/0000-0002-3008-8088}{\includegraphics[scale=0.06]{orcid.pdf}\hspace{1mm}Domniki Asimaki} \\
        Division of Engineering and Applied Sciences\\
        California Institute of Technology\\
        Pasadena, California, U.S.A. \\
    	\texttt{domniki@caltech.edu} \\ 
        \And
        \href{https://orcid.org/0000-0003-4729-6884}{\includegraphics[scale=0.06]{orcid.pdf}\hspace{1mm}Trevor J Carey} \\ Department of Civil \& Mineral Engineering\\
        The University of Toronto\\
        Toronto, ON, Canada\\
        \texttt{trevor.carey@utoronto.ca}
        \And
        \href{https://orcid.org/0000-0003-4279-2854}{\includegraphics[scale=0.06]{orcid.pdf}\hspace{1mm}Henry Benjamin Mason} \\
        University of Nevada, Reno\\
        College of Science\\
        Department of Geological Sciences and Engineering\\
        Reno, Nevada, U.S.A.\\
        \texttt{hbmason@unr.edu}
        } 
\fi
\if\doctype1
    \author{Grigorios Lavrentiadis \affilnum{1,2}, meeri,  Domniki Asimaki\affilnum{1}, meeri, and H Benjamin Mason \affilnum{2}, meeri, Trevor\affilnum{3}, meeri }
    \affiliation{
        \affilnum{1}California Institute of Technology, Pasadena, California\\
        \affilnum{2}State University of New York at Buffalo, Buffalo, New York \\
        \affilnum{3}University of Nevada, Reno, Reno, Nevada \\
        \affilnum{4}University of Toronto, Canada\\
    }
    \corrauth{Grigorios Lavrentiadis, Mechanical and Civil Engineering, 1200 E California blvd, MC 104-44, Pasadena, CA 91125}
    \email{glavrnet@caltech.edu}
\fi
\ifnum\doctype>1
  \ifnum\doctype<5
    \author[*1,2\orc{0000-0001-6546-1340}]{Grigorios Lavrentiadis}%
    \author[1\orc{0000-0002-3008-8088}]{Domniki Asimaki}
    \author[3\orc{0000-0003-4729-6884}]{Trevor J Carey}
    \author[4\orc{0000-0003-4279-2854}]{Henry Benjamin Mason}
    
    \affil[1]{First Affiliation, Institute, City, Country}{\auorc{0000-0000-0000-0001}{(AAF)}\auorc{0000-0000-0000-0002}{(ABS)}\auorc{0000-0000-0000-0003}{(ACT)}\auorc{0000-0000-0000-0004}{(AF)}}
    \affil[2]{Second Affiliation, Institute, City, Country}{\auorc{0000-0000-0000-0005}{(ADF)}\auorc{0000-0000-0000-0006}{(AS)}}
    \affil[3]{Third Affiliation, Institute, City, Country}{}
    \corau{*Corresponding author: glavrent@caltech.edu}
  \fi
\fi

\ifnum\flagreview>0
    \usepackage{blindtext}
    \usepackage{setspace}
\fi

\begin{document}

\if\doctype0
    \maketitle
\fi
\begin{abstract} 
    Accurately capturing near-fault ground motion characteristics, such as permanent tectonic offsets and dynamic velocity pulses induced by rupture directivity, among other effects, is crucial for seismic hazard assessment and engineering design.  This study presents an extension of the baseline correction method by \citet{Boore2001EffectEarthquake, Boore2002CommentsEarthquake} synchronously processing all three strong-motion components to preserve near-fault phenomena. The proposed approach is applied to the recordings of \mbox{M\,6.5} Guanshan foreshock and \mbox{M\,6.9} Chihshang mainshock from the near-fault stations ($R_{rup} < 20 km$), significantly increasing the number of ground motions containing near-fault effects.
    Furthermore, a validation with high-rate GNSS data demonstrates that the proposed approach is able to accurately capture the amplitude of tectonic offset and direction. 
    Structural damage observed during the reconnaissance visit strongly correlates with the locations of forward directivity (high peak-to-peak velocities in the fault normal direction) and tectonic offset in the fault parallel and vertical direction, further showcasing the significance of near-field effects in structural performance.  
    A simple classification scheme is also introduced to identify whether a record contains a fling-step pulse, a dynamic velocity pulse, or neither. Analysis of the extracted velocity pulses shows that fling-step pulses tend to occur early in the ground motion, while dynamic pulses span a broader temporal range. 
\end{abstract}
\ifnum\doctype>0
    \maketitle
\fi

\section{Introduction}

\ifnum\flagreview>0
    \large
    \linenumbers
\fi

Taiwan's high seismicity and dense network of high-quality strong-motion instruments have made it a cornerstone for advancing seismology, seismic hazard assessment, and geotechnical and structural engineering research and practice. The impetus was the 1999 \mbox{M,7.6} Chi-Chi earthquake, which highlighted the significance of near-field ground motion effects, including rupture directivity and hanging-wall amplification \citep{Shin2001, Yu2001, Shabestari2003, Chang2004}, as well as their impact on the performance of both geotechnical and structural systems \citep{Uzarski1999ChiChi, Abrahamson1999GEER002, Naeim2000, Loh2002}.

Motivated by the same goals, a reconnaissance mission was conducted in October 2022, following the \mbox{M,6.5} Guanshan foreshock (September 17, 2022, at 21:41:19.1 China Standard Time; GMT+08:00) and the \mbox{M,6.9} Chishang mainshock (September 18, 2022, at 14:44:15.2 CST). This effort, supported by the GeoEngineering Extreme Events Reconnaissance (GEER) Association and joined by scientists from Academia Sinica and the National Center for Research on Earthquake Engineering (NCREE), aimed to collect perishable field data. Due to the travel restrictions imposed by the COVID-19 pandemic, which delayed the response, the team focused on structural and geotechnical systems, where post-earthquake evidence remained visible. To understand their behavior, especially in the near-fault region, we obtained and processed the available ground motion records. During this process, it became evident that improvements to baseline correction procedures were necessary to robustly preserve near-fault features such as velocity pulses due to rupture directivity and fling-step effects.

Preserving near-fault effects in accelerograms is essential for assessing the seismic response of infrastructure near active faults, because features such as permanent (fling-step) displacement and rupture directivity have been repeatedly linked to damage in tall, long-period, and spatially distributed systems \citep[e.g.,][]{Loh2002, Chen2023, Kazaz2024}. Taiwan’s dense strong-motion networks have been central to revealing near-fault effects and their engineering consequences \citep{Shin2001, Yu2001, Shabestari2003, Chang2004, Uzarski1999ChiChi, Abrahamson1999GEER002, Naeim2000}. However, research targeted at the effects of near-fault ground motions on infrastructure remains limited by data scarcity as well as by processing practices that can suppress near-fault phenomena.

Raw acceleration records often contain spurious baseline offsets that, while negligible for acceleration-based measures, can induce substantial integration drift in the derived velocity and displacement records. Such offsets may arise from sensor tilt, transducer hysteresis, and related instrumental effects \citep{Graizer1979, Graizer2010, Iwan1985SomeAccelerograph}. 
Over the past decades, component-wise baseline-correction procedures have been developed to mitigate integration drift while preserving the near-field characteristics of strong ground motions \citep{Boore2001EffectEarthquake, Boore2002CommentsEarthquake}. A key limitation of earlier approaches is the identification of the portion of the signal affected by spurious artifacts. More recent studies have sought to better formalize and automate this selection process \citep{Wang2011, DAmico2019}. 
Despite these advances, robust identification of the affected signal segment on a component-by-component basis remains challenging, as multiple plausible interval selections may yield visually acceptable corrected records while implying different permanent displacements. 
To address this limitation, we propose a baseline-correction framework that builds upon existing methodologies but enforces consistent selection of correction intervals across all three ground-motion components, which enables cross-component validation of inferred tectonic displacements and improves robustness in single-component processing.

The 2022 Chishang earthquake sequence provides an ideal testbed for evaluating the seismic response of near-fault infrastructure. The event is characterized by dense near-fault strong-motion recordings, independently measured co-seismic displacements from Global Navigation Satellite System (GNSS) stations for validation, and well-documented bridge performance data collected during our post-earthquake reconnaissance. 
In this study, we (i) demonstrate the proposed baseline-correction methodology, (ii) validate the inferred permanent displacements against independent GNSS observations, (iii) apply the procedure to ground motions recorded in Taiwan’s Longitudinal Valley during the 2022 Chishang sequence, and (iv) relate the corrected near-fault response characteristics to observed bridge damage. Application of the method reveals pronounced, rupture-consistent velocity pulses and permanent ground displacements in the vicinity of damaged bridges, in the absence of liquefaction, highlighting the importance of preserving near-field phenomena in both analysis and design. 
Beyond the methodological contributions and reconnaissance findings, we provide a comprehensive, openly accessible near-fault ground-motion dataset, including processed time histories and intensity measures, to be used by the broader science and engineering community in advancing the understanding and modeling of near-field effects.

\section{Strong-Motion Dataset}\label{sec:data}

A total of 35 strong-motion records from the \mbox{M\,6.5} Guanshan earthquake and 24 from the \mbox{M\,6.9} Chishang earthquake were obtained from near-fault instruments operated by the Taiwan Strong Motion Instrumentation Program (TSMIP), the Earthquake Early Warning System (EEWS), and the Central Weather Bureau Seismic Network (CWBSN). Records from co-located stations with instruments affected by clipping or malfunction, or those with missing portions of the events, were excluded from the final dataset. 
A set of GNSS stations that captured the total permanent deformation and were adjacent to the strong-motion stations was also used to validate the proposed baseline correction method. Table~\ref{tab:smsn_recordings} summarizes the strong-motion stations that recorded each event, the corresponding nearby GNSS station, and the distance between the two instruments.

\begin{figure}[htbp!]
    \centering
    \includegraphics[width=0.95\textwidth]{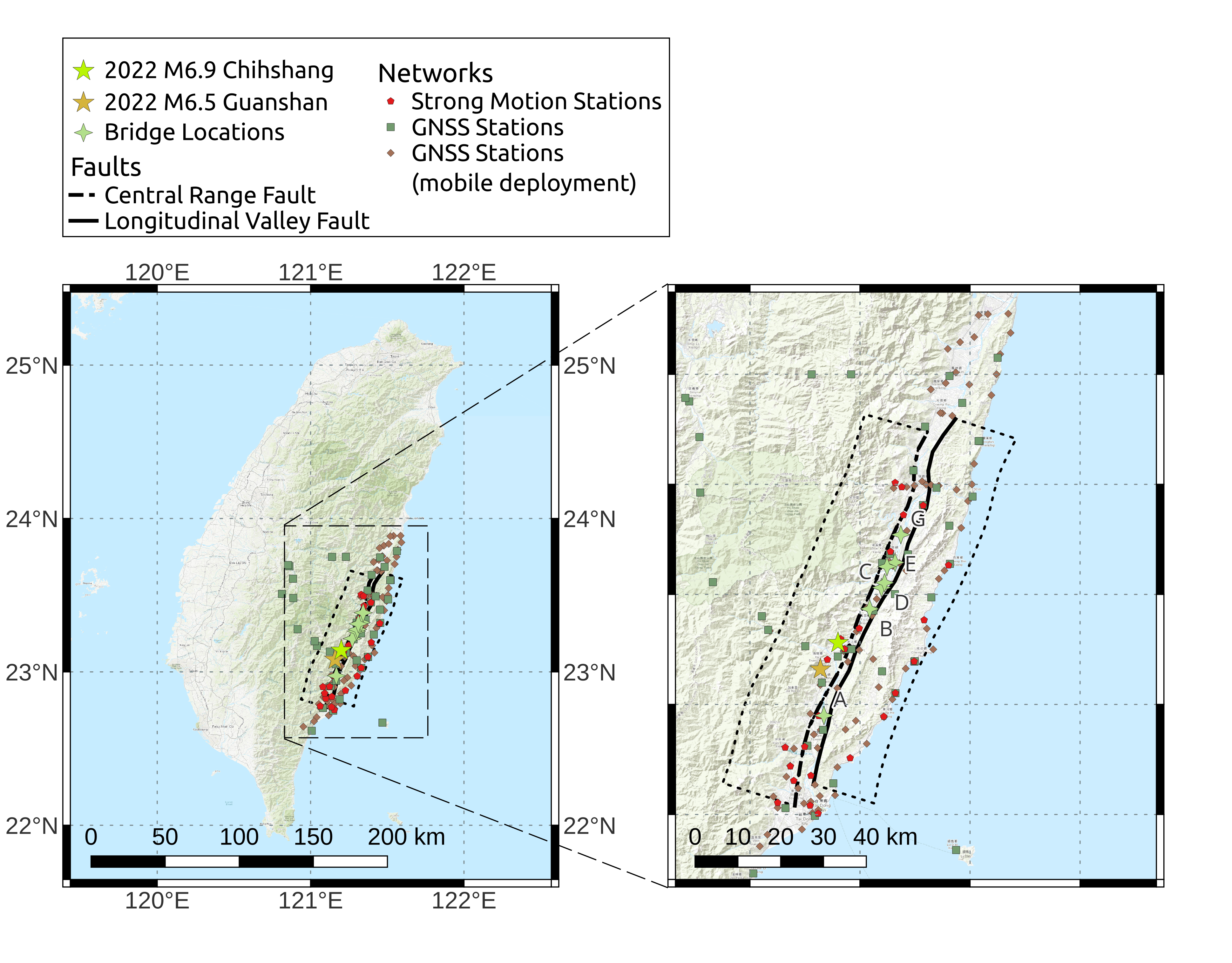}
    \caption{Event and station locations for the $M 6.9$ Chishang and $M 6.5$ Guanshan earthquakes. The stars represent the epicenter locations.}
    \label{fig:eq_sta_location}
\end{figure}

Figure~\ref{fig:eq_sta_location} illustrates the spatial distribution of the selected strong-motion stations together with the locations of the bridges surveyed during the post-earthquake reconnaissance. Table~\ref{tab:bridgedamage} summarizes the surveyed bridges (also shown in Figure~\ref{fig:eq_sta_location}), including their geographic coordinates (latitude and longitude), bridge orientation (azimuth), the nearest strong-motion station, and the corresponding bridge-to-station distance.

\begin{figure}[htbp!]
    \centering
    \includegraphics[width=0.5\textwidth]{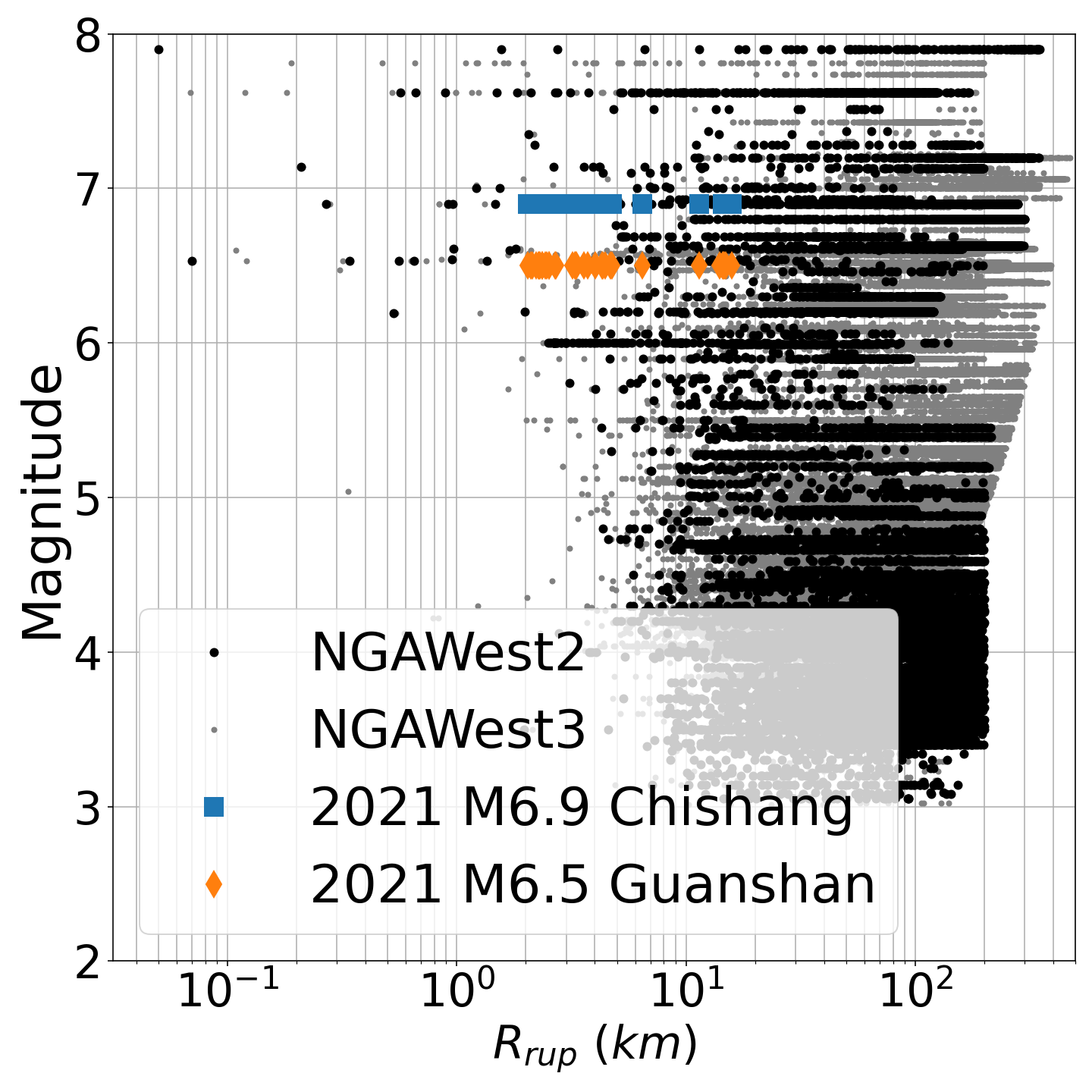}
    \caption{Comparison of magnitude-distance distribution for the $M 6.9$ Chishang and $M 6.5$ Guanshan earthquakes with the NGA-West2 dataset}
    \label{fig:data_distribution}
\end{figure}

Figure~\ref{fig:data_distribution} compares the magnitude–distance distributions of the Guanshan and Chishang ground motions with those in the NGA-West2 database \citep{Ancheta2014} and NGA-West3 database \citep{Stewart2024}.
Of the selected recordings, 46 were obtained at rupture distances less than 10~km, which is equivalent to approximately 35\% of all \mbox{$R_{\mathrm{rup}} < 10$~km} ground motions from earthquakes with magnitudes greater than 6.5 in the NGA-West2 dataset, and 22\% of the NGA-West3 dataset, not including the ground motion recordings from this sequence; thereby significantly expanding the empirical database of near-fault motions exhibiting forward directivity and fling-step effects. 
Furthermore, ground motions in the NGAWest2 and NGAWest3 datasets, including the near-fault records and the records of this sequence, were baseline corrected through high-pass filtering \citep{Ramos2023}, distorting or erasing key near-fault characteristics such as velocity pulses and permanent displacement offsets. For TSMIP stations, site parameters such as $V_{S30}$ and $Z_{1.0}$ were obtained from the NCEE database. For stations operated by other networks, the parameters from NGAWest3 were adopted. 
Rupture distance and other distance metrics were computed using the finite-fault geometry provided by \citet{Lee2023}.

\begin{table}[htbp!]
    \centering
    \caption{Summary of strong-motion and GNSS station recordings for the 2022 $M 6.9$ Chihshang and $M 6.5$ Guanshan earthquakes. The table lists station locations, recording status, adjacent GNSS station (permanent and mobile deployments), and the corresponding distances between the instruments.}
    \begin{tabular}{l c c|c c c|c c c}
    \hline
    \multirow{2}{*}{Station} & Latitude & Longitude & \multicolumn{3}{c|}{$M 6.9$ Chihshang} &  \multicolumn{3}{c}{$M 6.5$ Guanshan} \\
     & (deg) & (deg) & Recorded & GNSS & \thead{$\Delta$ (km)} & Recorded & GNSS & \thead{$\Delta$ (km)} \\
    \hline
    A330 & 22.827 & 121.100 & $\checkmark$ & E306 (mobile) & 2.12 &  &  &  \\
    EHY & 23.504 & 121.330 & $\checkmark$ & GA35 (mobile) & 1.42 & $\checkmark$ & JPEI & 5.24 \\
    HWA004 & 23.173 & 121.248 & $\checkmark$ &  &  & $\checkmark$ &  &  \\
    HWA036 & 23.498 & 121.374 &  &  &  & $\checkmark$ & JPEI & 3.78 \\
    HWA037 & 23.452 & 121.394 & $\checkmark$ &  &  & $\checkmark$ & CHUN & 0.11 \\
    HWA039 & 23.383 & 121.359 &  &  &  & $\checkmark$ & JPIN & 4.66 \\
    HWA041 & 23.265 & 121.303 &  &  &  & $\checkmark$ & FUDN & 3.15 \\
    HWA042 & 23.220 & 121.264 &  &  &  & $\checkmark$ &  &  \\
    HWA054 & 23.430 & 121.349 & $\checkmark$ &  &  & $\checkmark$ &  &  \\
    HWA073 & 23.347 & 121.319 & $\checkmark$ & E801 (mobile) & 0.63 & $\checkmark$ & JULI & 0.62 \\
    HWA075 & 23.494 & 121.345 & $\checkmark$ & GA34 (mobile) & 1.13 & $\checkmark$ & JPEI & 4.97 \\
    S007 & 23.417 & 121.370 &  &  &  & $\checkmark$ &  &  \\
    S027 & 23.117 & 121.169 &  &  &  & $\checkmark$ &  &  \\
    S047 & 23.284 & 121.286 &  &  &  & $\checkmark$ &  &  \\
    S054 & 22.770 & 121.137 & $\checkmark$ & 934 (mobile)  & 0.99 & $\checkmark$ &  &  \\
    S055 & 22.904 & 121.124 & $\checkmark$ & LONT & 0.69 & $\checkmark$ & LONT & 0.69 \\
    TTN001 & 23.316 & 121.451 & $\checkmark$ & PING & 0.48 & $\checkmark$ & PING & 0.48 \\
    TTN002 & 22.973 & 121.304 & $\checkmark$ & S220 (mobile) & 0.15 & $\checkmark$ &  &  \\
    TTN014 & 23.098 & 121.373 & $\checkmark$ & CHEN & 0.07 & $\checkmark$ & CHEN & 0.07 \\
    TTN015 & 22.752 & 121.155 & $\checkmark$ & S264 (mobile) & 0.89 & $\checkmark$ &  &  \\
    TTN020 & 23.126 & 121.215 & $\checkmark$ & GC35 (mobile) & 0.12 & $\checkmark$ & TTCS & 0.12 \\
    TTN021 & 23.102 & 121.176 & $\checkmark$ & SHAN & 2.50 & $\checkmark$ & SHAN & 2.50 \\
    TTN022 & 23.095 & 121.219 &  &  &  & $\checkmark$ &  &  \\
    TTN023 & 23.051 & 121.164 &  &  &  & $\checkmark$ & KUA2 & 0.18 \\
    TTN025 & 22.902 & 121.080 & $\checkmark$ & 9176 (mobile) & 2.10 & $\checkmark$ &  &  \\
    TTN026 & 22.860 & 121.092 & $\checkmark$ & GE43 (mobile) & 0.11 & $\checkmark$ &  &  \\
    TTN028 & 22.777 & 121.063 & $\checkmark$ & S273 (mobile) & 1.30 & $\checkmark$ & TTUN & 2.30 \\
    TTN032 & 23.245 & 121.414 &  &  &  & $\checkmark$ & NIPU & 0.26 \\
    TTN033 & 23.192 & 121.396 & $\checkmark$ & A056 (mobile) & 2.30 & $\checkmark$ &  &  \\
    TTN035 & 22.878 & 121.228 & $\checkmark$ & A081 (mobile) & 3.80 & $\checkmark$ &  &  \\
    TTN045 & 22.974 & 121.156 & $\checkmark$ &  &  & $\checkmark$ &  &  \\
    TTN047 & 22.838 & 121.139 & $\checkmark$ &  &  & $\checkmark$ &  &  \\
    TTN048 & 22.772 & 121.091 &  &  &  & $\checkmark$ & TTUN & 1.30 \\
    TTN057 & 23.026 & 121.330 & $\checkmark$ & DULI & 0.02 & $\checkmark$ & DULI & 0.02 \\
    TTN061 & 23.149 & 121.206 & $\checkmark$ &  &  & $\checkmark$ &  &  \\
    \hline
    \end{tabular}
    \label{tab:smsn_recordings}
\end{table}

\begin{table}[htbp!]
    \centering
    \caption{Bridges surveyed during post-earthquake reconnaissance following the 2022 Chishang Earthquake ($\Delta$ is nearest station distance, and the azimuths are computed using the locations of the bridge abutments).}
    \begin{tabular}{ c c | c c c c c}
    \hline
    \thead{Label} & \thead{Name} & \thead{Latitude} & \thead{Longitude} & \thead{Azimuth\\(deg)} & \thead{Nearest\\Station} & \thead{$\Delta$\\(km)}  \\
    \hline
     A & Baohua     & 22.975 & 121.168 & 93  & TTN045 & 1.20 \\
     B & Luntain    & 23.217 & 121.272 & 118 & HWA004 & 5.50 \\
     C & Zhangfu    & 23.265 & 121.299 & 121 & HWA073 & 9.32 \\
     D & Dongli RR  & 23.277 & 121.306 & 23  & HWA073 & 7.97 \\
     E & Yuli (2X)  & 23.323 & 121.330 & 102 & HWA073 & 2.92 \\
     F & Kecheng RR & 23.311 & 121.312 & 356 & HWA073 & 3.87 \\
     G & Gaoliao    & 23.386 & 121.343 & 99  & HWA073 & 4.91 \\
    \hline
    \end{tabular}
    \label{tab:bridgedamage}
\end{table}

\section{Methodology: Baseline Processing} \label{sec:methodology}

\begin{figure}[htbp!]
    \begin{subfigure}[t]{0.32\textwidth}
        \caption{}
        \includegraphics[height = 0.95\textwidth]{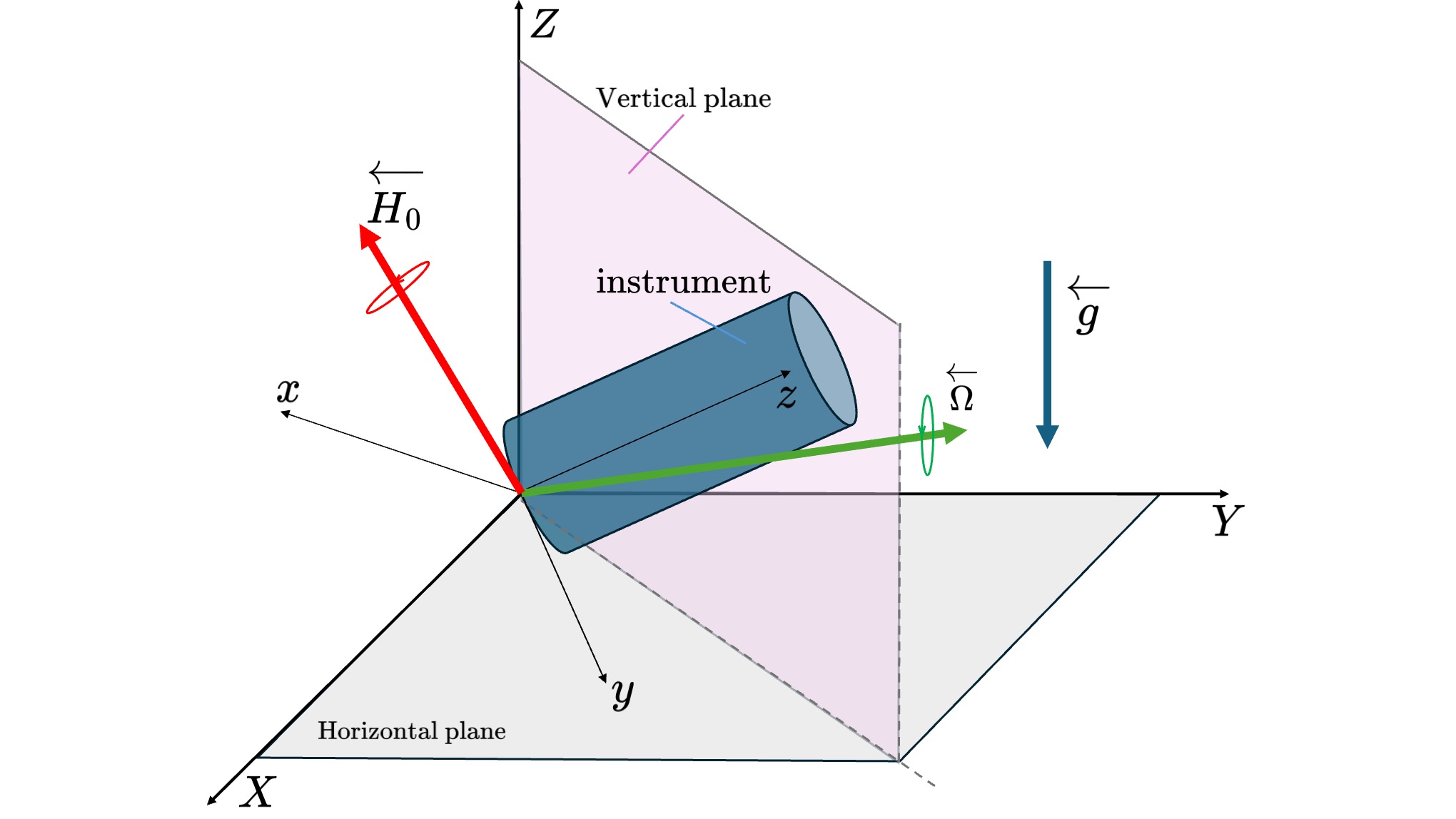}
    \end{subfigure} 
    \hspace{0.1\textwidth}
    \begin{subfigure}[t]{0.32\textwidth}
        \caption{}
        \includegraphics[height = 0.95\textwidth]{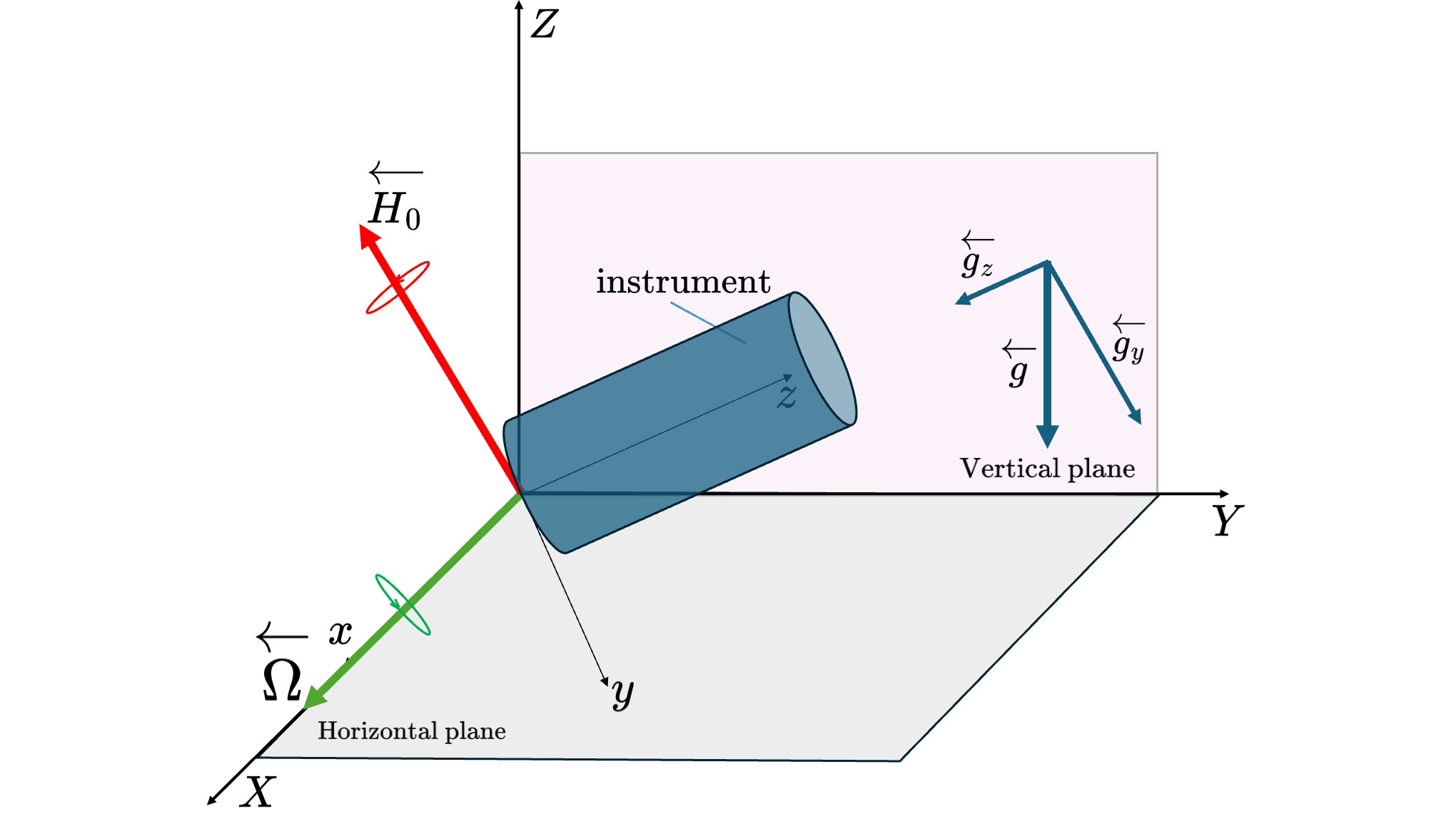}
    \end{subfigure}  
    \caption{Instrument tilt (a) about an arbitrary axis, (b) about the X axis.}
    \label{fig:instrument}
\end{figure}

A strong-motion instrument can be idealized as a set of three mass–spring–damper systems (one for each direction of motion), mounted inside a nearly rigid instrument casing. The casing, in turn, is attached to the ground surface, whose accelerations are to be determined. If we model the casing as a rotating body about a fixed point (the point of contact of the casing to the ground), then the instrument as a whole is required to satisfy the following two vector equations in three dimensions:

\begin{align}
    \sum \bf{F} = m ~ \overline{\bf{a}}\\
    \sum \bf{M_0} = \bf{\dot{H}}_{0}
\end{align}

namely Newton's second law and the conservation of angular momentum, where $\sum \bf{F}$ is the sum of external forces and $\sum \bf{M_0}$ the sum of external moments with respect to the fixed point of rotation, $0$, applied to the instrument. These equations are valid in the most general case of rigid body motion. If the instrument is subjected to a sudden tilt about an arbitrary axis, idealized here by the angular velocity vector $\bf{\Omega}$ in Figure~\ref{fig:instrument}, then the equations of motion for each of the three transducers in the instrument would be formulated as follows:

\begin{align}
    \text{x direction:\qquad} & \ddot{u}_x(t) + 2 \zeta_x \omega_x \dot{u}_x(t) + \omega_x^2 u_x(t) = -\ddot{u}_{g,x}(t) + f(\overline{I}, \bf{\Omega}, \bf{\dot{\Omega}}, \bf{g}) \\
    \text{y direction:\qquad} & \ddot{u}_y(t) + 2 \zeta_y \omega_y \dot{u}_y(t) + \omega_y^2 u_y(t) = -\ddot{u}_{g,y}(t) + f(\overline{I}, \bf{\Omega}, \bf{\dot{\Omega}}, \bf{g})   \\
    \text{z direction:\qquad} & \ddot{u}_z(t) + 2 \zeta_z \omega_z \dot{u}_z(t) + \omega_z^2 u_z(t) = -\ddot{u}_{g,z}(t) + f(\overline{I}, \bf{\Omega}, \bf{\dot{\Omega}}, \bf{g})  
\end{align}

where $u_x(t)$, $u_y(t)$, and $u_z(t)$ denote the oscillator's relative displacement in the horizontal ($x$, $y$) and vertical ($z$) directions, respectively, with dots representing time derivatives (i.e., single dot for velocity and double dot for acceleration). The term $\omega$ refers to each transducer's natural frequency, $\zeta$ to the fraction of critical damping of each transducer's dashpot, and $\ddot{u_g}$ to the ground (base) acceleration. The function $f(\overline{I}, \bf{\Omega}, \bf{\dot{\Omega}}, \bf{g})$ emerges from Newton's second law and the conservation of angular momentum as the instrument tilts, where $\overline{I}$ is the inertia tensor of the instrument casing relative to the point of rotation, $\bf{\Omega}$ is the angular velocity vector, $\bf{\dot{\Omega}}$ is the derivative of the angular velocity vector (not to be confused with the angular acceleration which is the simplified case in 1D) and $\bf{g}$ is the acceleration of gravity vector.

In the special case that the rotation occurs only about one axis (e.g. axis X: ${\bf \Omega} = \Omega_x \hat{i}$ as shown in Figure \ref{fig:instrument}b) and the instrument is symmetric with respect to the YZ plane (in which case the products of inertia $I_{xy}$ and $I_{xz}$ are zero), equations (1) and (2) would be simplified as follows:

\begin{align}
    \sum M_x = I_x ~ \dot{\Omega}\\
    \sum F_y = m ~ g_y\\
    \sum F_z = m ~ g_z
\end{align}

where $\sum M_x$ is the sum of moments about the $x$-axis of rotation, and $\sum F_y$ and $\sum F_z$ are the sums of forces in the directions normal to the rotation axis ($y$-direction) and in the vertical ($z$-direction), respectively.

From these equations, it follows that even if the instrument rotates about one axis (here, $x = X$), the rigid body motion of the casing introduces spurious acceleration offsets $g_y$ and $g_z$ in the $y$ and $z$ directions, to satisfy equilibrium and conserve angular momentum. These phenomena are coupled and synchronous, and thus correcting the acceleration time series to remove the spurious effects of instrument tilt also needs to be accounted for in a coupled and synchronous approach.


Following these observations, the \cite{Boore2001EffectEarthquake, Boore2002CommentsEarthquake} baseline correction method, as well as the timing selection criteria of \cite{Wang2011}, are adopted to remove spurious acceleration offsets from the three components by selecting the same time instances for the baseline algorithm for all three components due to the commonality of the tilt angles. 
The proposed procedure comprises two stages. First, the acceleration time histories are corrected for pre--P-wave offsets. Second, the time intervals used in the baseline algorithm are iteratively optimized until a physically consistent fit is achieved simultaneously for all three components. If the automatic optimization fails to converge, the correction intervals are selected manually, subject to a visual inspection that the resulting velocity and displacement traces remain plausible across components.

The two implemented baseline correction methods are the linear and quadratic approaches applied to the velocity time histories. 
The linear method is designed to capture sudden changes in instrument tilt, which introduce a step-like (constant) acceleration offset, resulting in linear trends in the velocity time history and corresponding quadratic trends in displacement.
In contrast, the quadratic method addresses gradual changes in instrument tilt, which produce linearly varying acceleration trends that manifest as quadratic trends in velocity and cubic trends in displacement.

\begin{figure}[htbp!]
    \centering
    \begin{subfigure}[t]{0.4\textwidth}
        \caption{}
        \includegraphics[width = \linewidth]{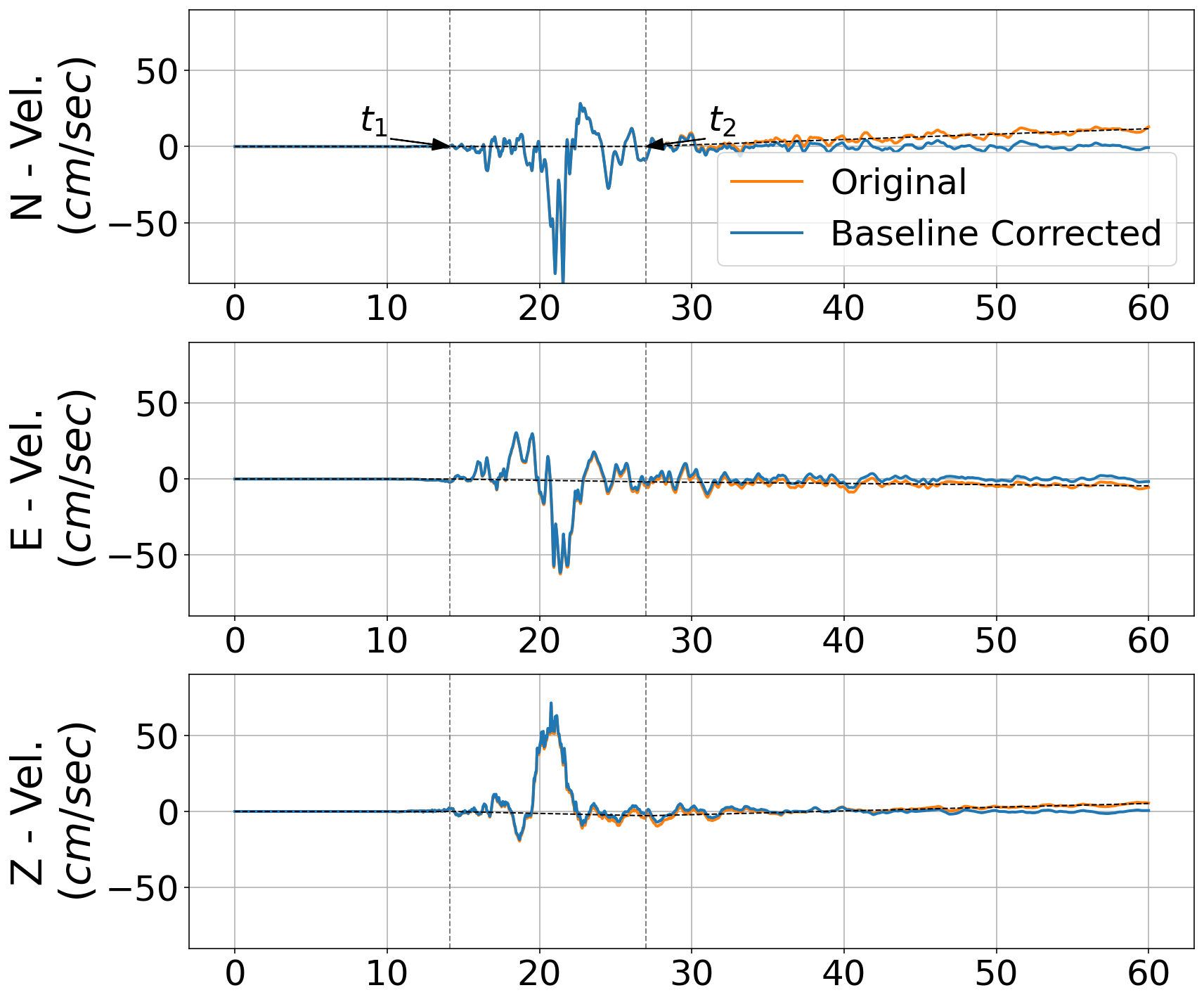}
    \end{subfigure} 
    \hspace{0.1\textwidth}
    \begin{subfigure}[t]{0.4\textwidth}
        \caption{}
        \includegraphics[width = \linewidth]{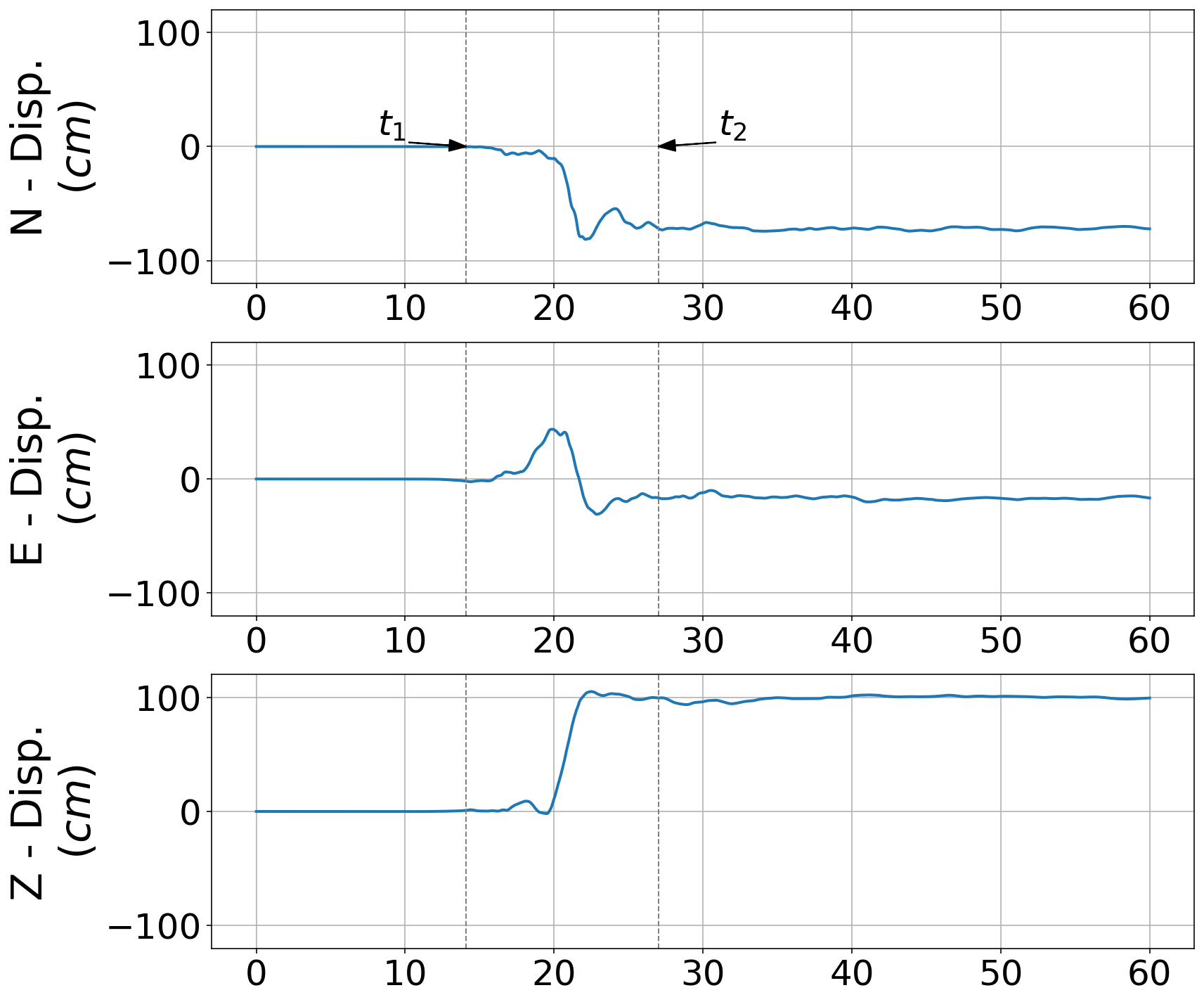}
    \end{subfigure}  
    \caption{Quadratic baseline correction applied to the three-component time history of the \mbox{M\,6.9} Chishang earthquake recorded at the HWA073 station ($R_{\mathrm{rup}} = 2.4$~km). 
    (a) Raw and corrected velocity time histories for all three components. 
    (b) Corrected displacement time histories, illustrating the removal of spurious trends and recovery of permanent displacements.}
    \label{fig:examp_piecewise_lienar}
\end{figure}

In the linear correction approach, two time instances, $t_1$ and $t_2$, are selected to correspond to sudden changes in acceleration offset. The final acceleration offset $a_f$, assumed constant after time $t_2$, introduces a spurious velocity of the form:
\begin{equation} \label{eq:v_f}
    v_f(t) = v_{f0} + a_f t \quad \text{for} \quad t \ge t_2
\end{equation}
where $v_{f0}$ is the initial velocity at time $t_2$. Recognizing that the corrected velocity time history should be zero at the end, $v_{f0}$ and $a_f$ are determined by fitting Equation \ref{eq:v_f} to the recorded velocity $v(t)$ using a least-squares loss:
\begin{equation} \label{eq:a_f}
    v_{f0}, a_f = \arg\min_{v_{f0}, a_f} \int_{t \ge t_2} \left[ v(t) - v_f(t) \right]^2 dt
\end{equation}
where $v(t)$ is the raw (uncorrected) velocity time history.
Between times $t_1$ and $t_2$, the acceleration offset $a_m$ introduces a linearly increasing spurious velocity given by:
\begin{equation} \label{eq:v_m}
    v_m(t) = a_m (t - t_1) \quad \text{for} \quad t_1 \le t \le t_2
\end{equation}
To ensure continuity in the corrected velocity at time $t_2$, the velocity functions must satisfy the condition $v_m(t_2) = v_f(t_2)$. Substituting from Equations~\ref{eq:v_f} and \ref{eq:v_m}, the acceleration offset $a_m$ is determined as:
\begin{equation}
    a_m = \frac{v_{f0} + a_f t_2}{t_2 - t_1}
\end{equation}
Figure~\ref{fig:examp_piecewise_lienar} illustrates the piecewise-linear baseline correction applied to the three-component time histories of the \mbox{M\,6.9} Chishang earthquake, recorded at station HWA073 at a rupture distance of $R_{\mathrm{rup}} = 2.4$~km. The corrected records demonstrate that the method was able to preserve the single-sided velocity pulse in the north and vertical components, which resulted in permanent displacement due to the fling effect, and a double-sided velocity pulse in the east component, indicative of rupture directivity.

\begin{figure}[htbp!]
    \centering
    \begin{subfigure}[t]{0.4\textwidth}
        \caption{}
        \includegraphics[width = \linewidth]{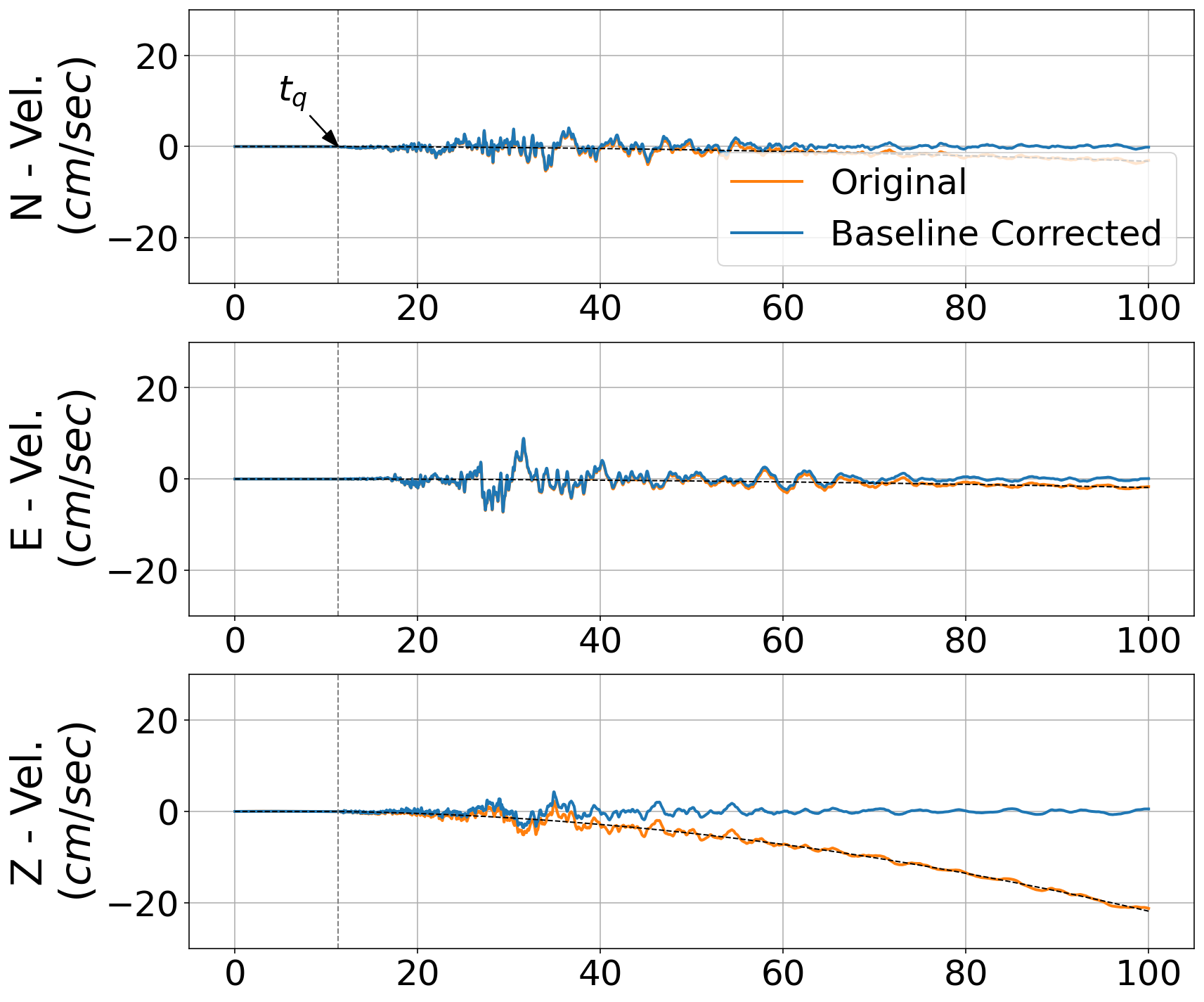}
    \end{subfigure} 
    \hspace{0.1\textwidth}
    \begin{subfigure}[t]{0.4\textwidth}
        \caption{}
        \includegraphics[width = \linewidth]{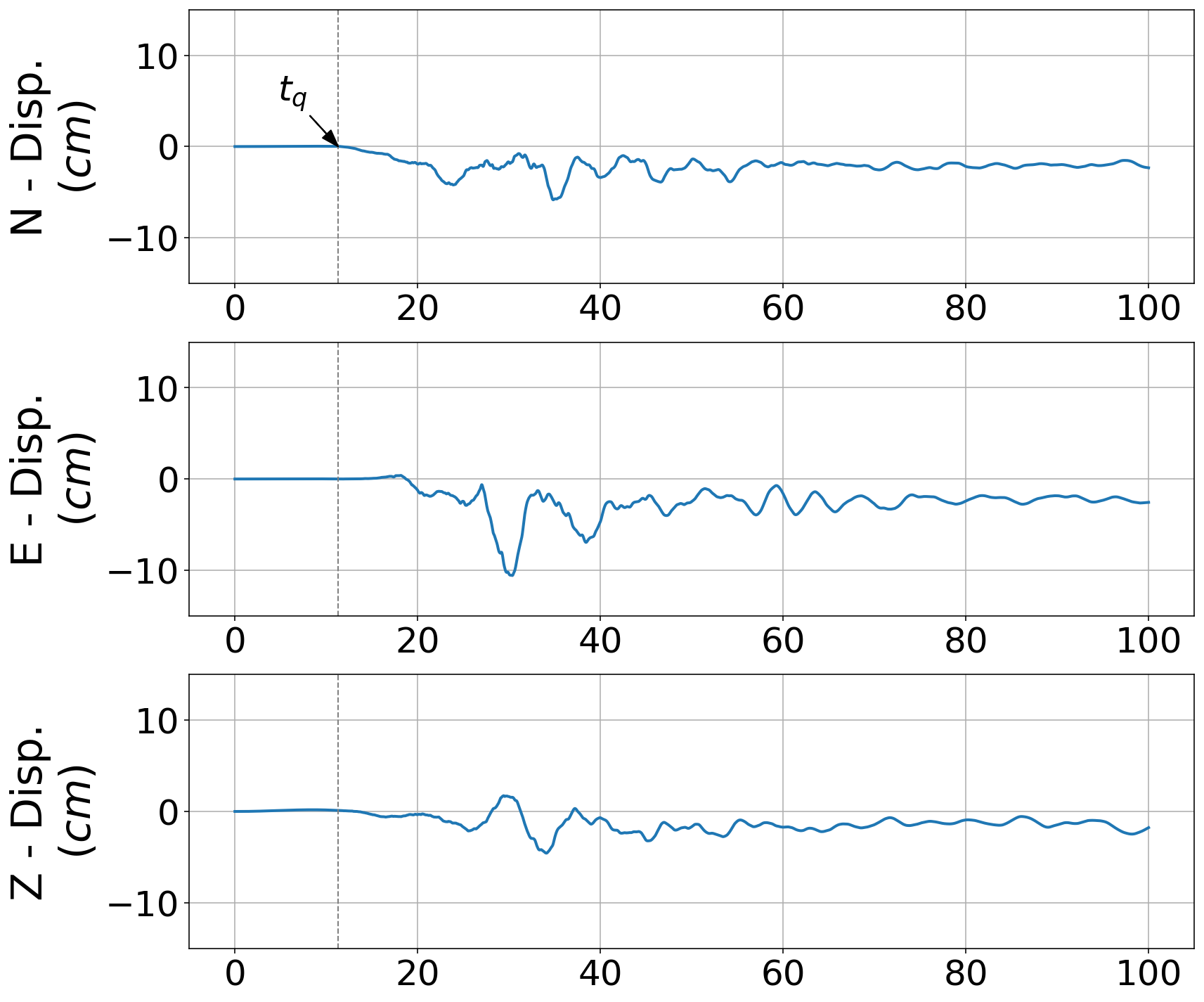}
    \end{subfigure}  
    \caption{Quadratic baseline correction applied to the three-component time history of the \mbox{M\,6.9} Chishang earthquake recorded at the TTN026 station ($R_{\mathrm{rup}} = 2.6$~km). 
    (a) Raw and corrected velocity time histories for all three components. 
    (b) Corrected displacement time histories, illustrating the removal of spurious trends and recovery of permanent displacements.}
    \label{fig:examp_quadratic}
\end{figure}

Alternatively, in the quadratic approach, a time instance $t_q$ is selected to define a linear trend in acceleration ($a_q \cdot t)$, which implies a quadratic spurious velocity of the form:
\begin{equation}
    v_q(t) = \frac{1}{2}  a_q  t^2 \quad \text{for} \quad t \ge t_q
\end{equation}
Similarly to the piecewise linear approach, the $a_q$ term is determined by fitting the raw time history with a least-squares loss. 
Figure \ref{fig:examp_quadratic} presents a quadratic baseline correction method for the \mbox{M\,6.9} Chishang earthquake time histories recorded at the station TTN026 at $R_{\mathrm{rup}} = 2.6$~km.

Key considerations for the correction of time histories are: which baseline method to use and how to select the time instances. 
If the velocity record extends long enough after the strong shaking, the first consideration is arguably easy to decide; it depends on the shape of the trend in the raw time histories. 
For instance, the time history recorded on the station TTN026 (Figure \ref{fig:examp_quadratic}) shows a clear quadratic velocity trend in the vertical component, indicating that the quadratic method is required. 
On the other hand, the HWA073 time history (Figure \ref{fig:examp_piecewise_lienar}) shows a linear trend after strong shaking, indicating that the piecewise linear method is more suitable. 


Selecting the appropriate time instances for baseline correction, on the other hand, is less straightforward and still requires some subjective judgment. In this study, for the piecewise-linear correction, $t_1$ and $t_2$ were selected before and after the strong shaking. 
For the quadratic correction, $t_q$ was chosen prior to the onset of strong shaking.
In both approaches, time selections were made using all three components of motion, as the onset of strong shaking is typically better identifiable in the vertical component, even when the horizontal components were of primary interest. 
In cases where any component exhibited a non-constant displacement trend after the strong shaking, the selected time instances were iteratively adjusted until satisfactory corrections were achieved across all three components.

\begin{figure}
    \centering
    \begin{minipage}{0.48\linewidth}
        \centering
        \begin{subfigure}{\linewidth}
          \centering
          \caption{}
          \includegraphics[width=0.9\linewidth]{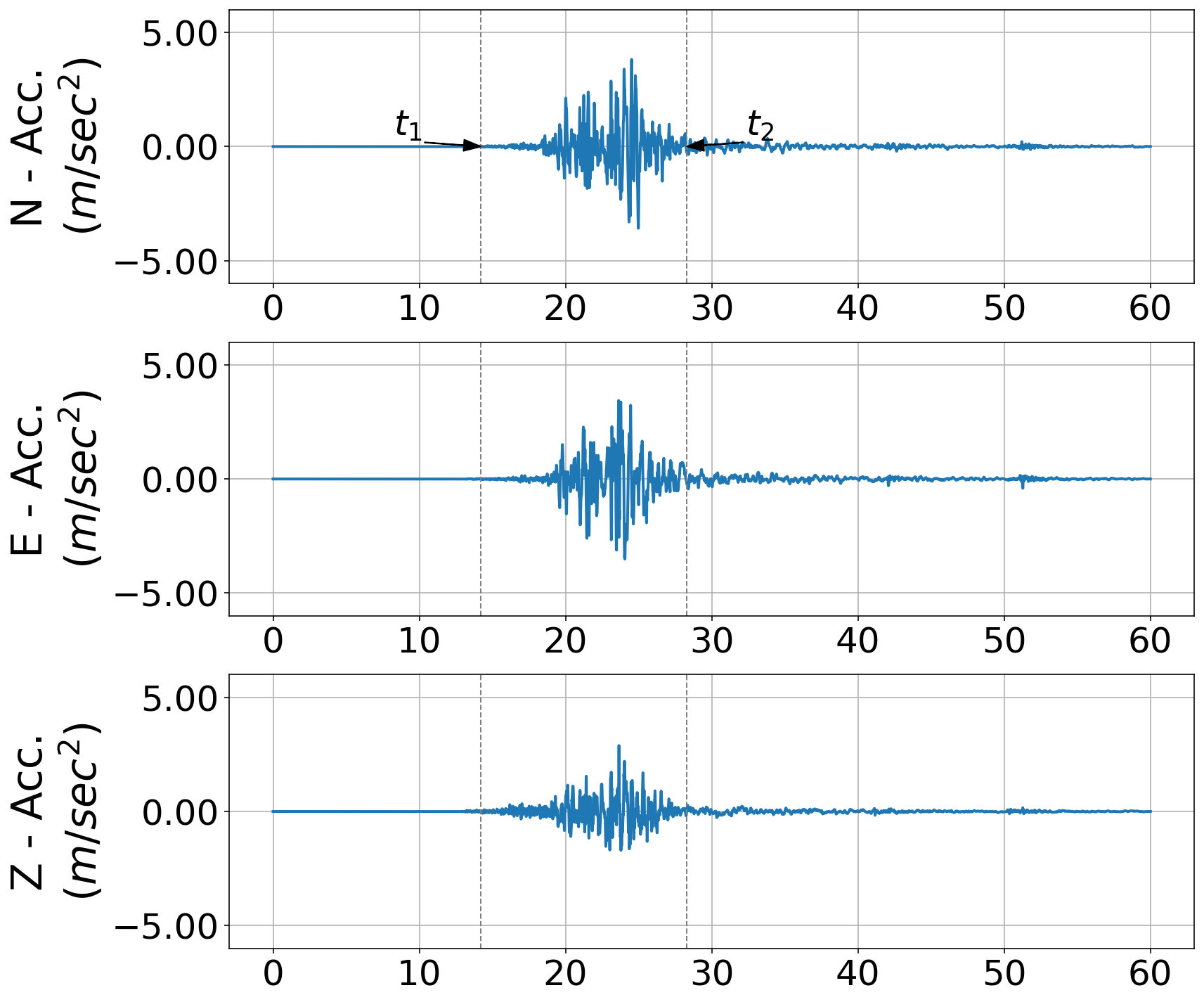}
        \end{subfigure}
        \begin{subfigure}{\linewidth}
          \centering
          \caption{}
          \includegraphics[width=0.9\linewidth]{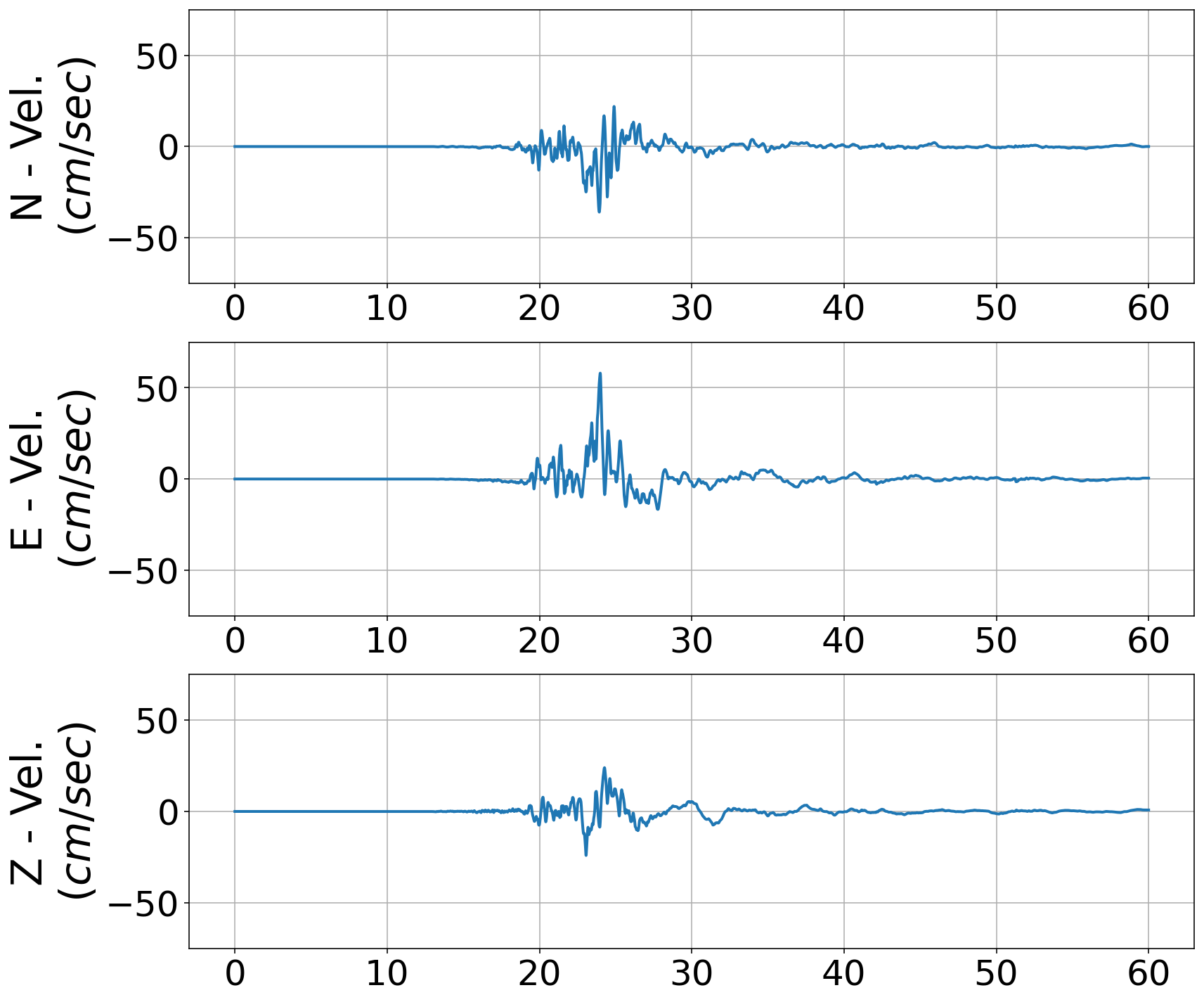}
        \end{subfigure}
        \begin{subfigure}{\linewidth}
          \centering
          \caption{}
          \includegraphics[width=0.9\linewidth]{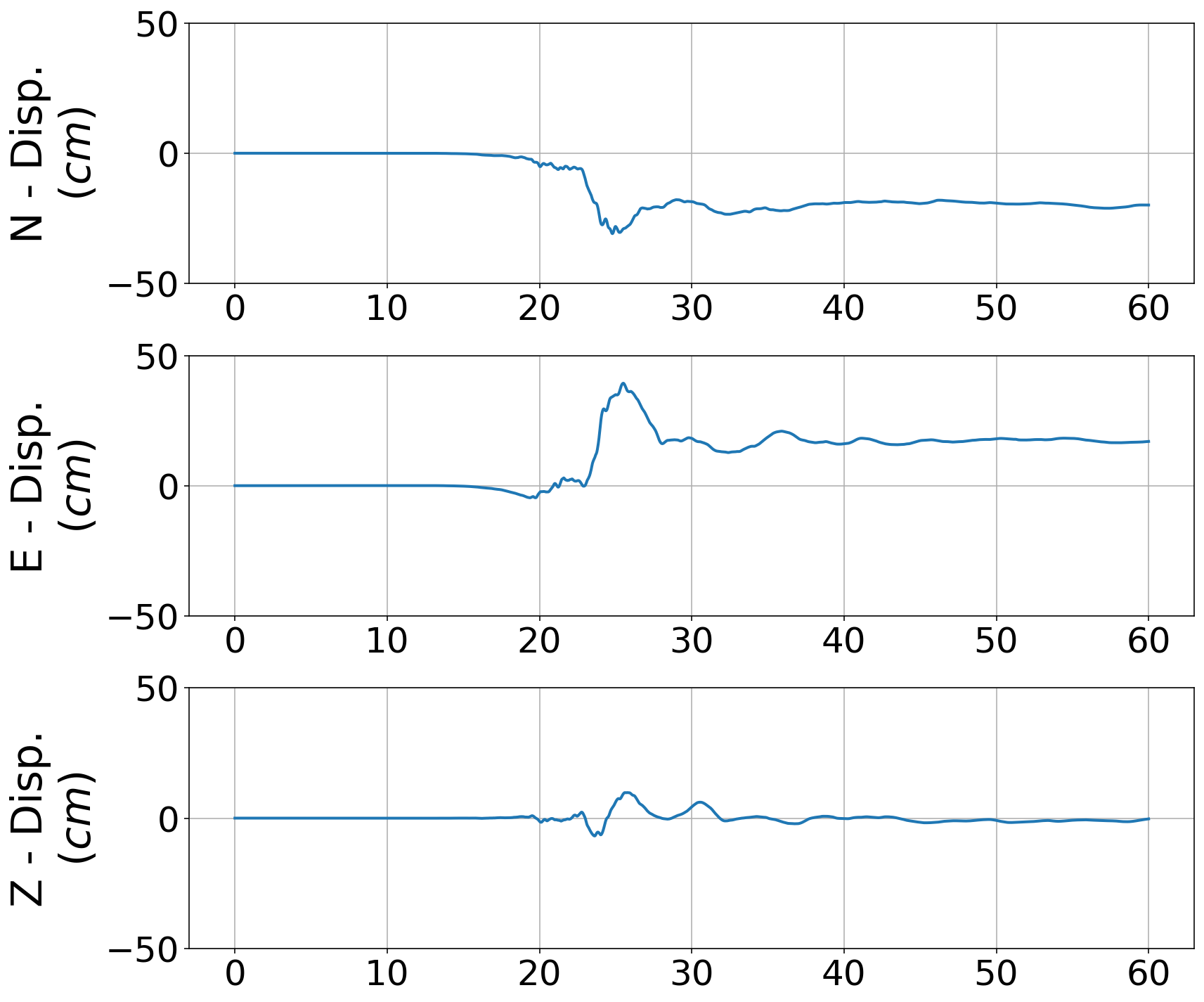}
        \end{subfigure}
        \caption{Baseline correction of ground motions recorded at station HWA075 during the \mbox{M\,6.9} Chishang earthquake, using timing selections based on all three components. (a) acceleration time histories, (b) velocity time histories, and (c) displacement time histories.}
        \label{fig:examp_timing_correct}
    \end{minipage}
    \hfill
    \begin{minipage}{0.48\linewidth}
        \centering
        \begin{subfigure}{\linewidth}
          \centering
          \caption{}
          \includegraphics[width=0.9\linewidth]{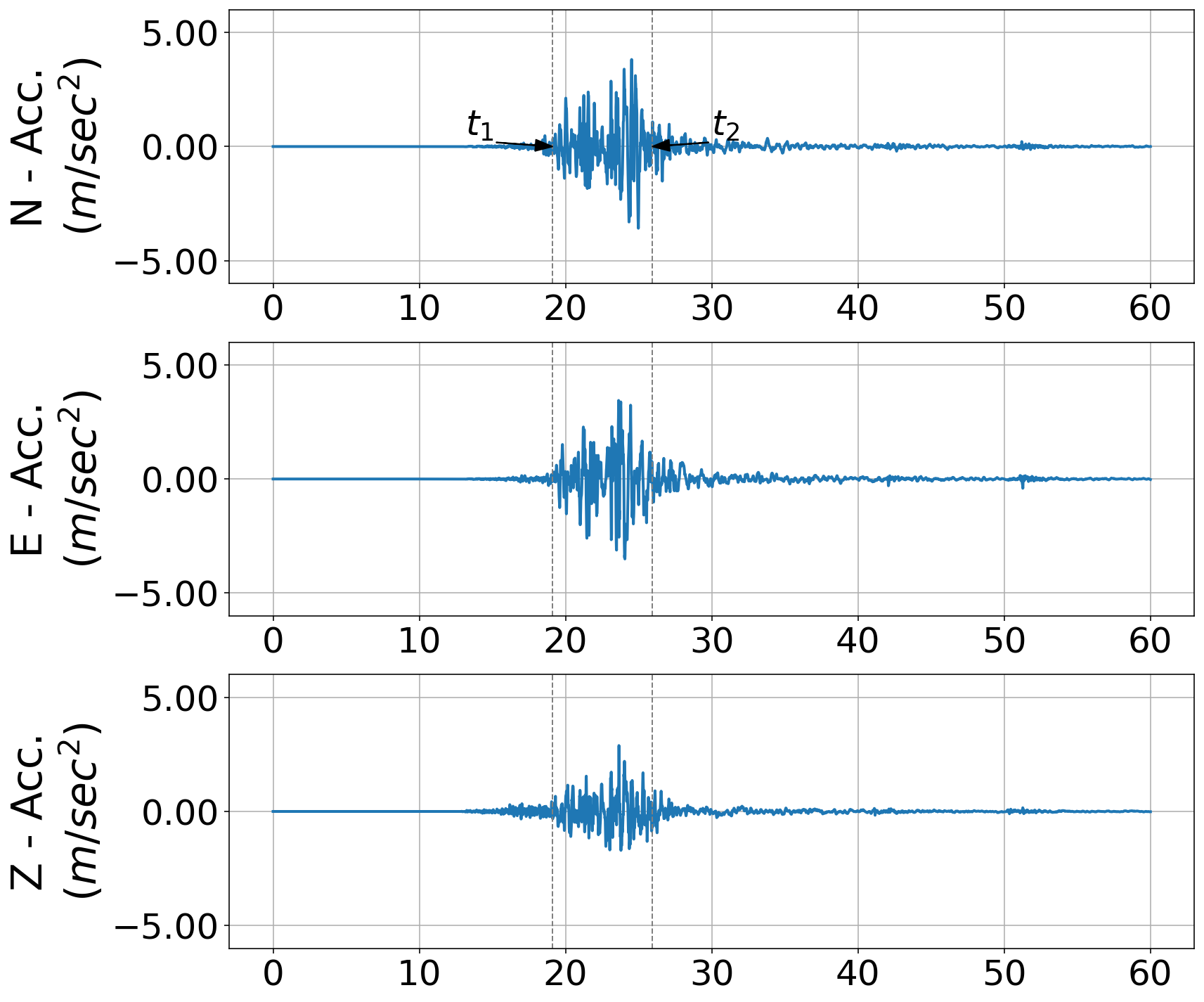}
        \end{subfigure}
        \begin{subfigure}{\linewidth}
          \centering
          \caption{}
          \includegraphics[width=0.9\linewidth]{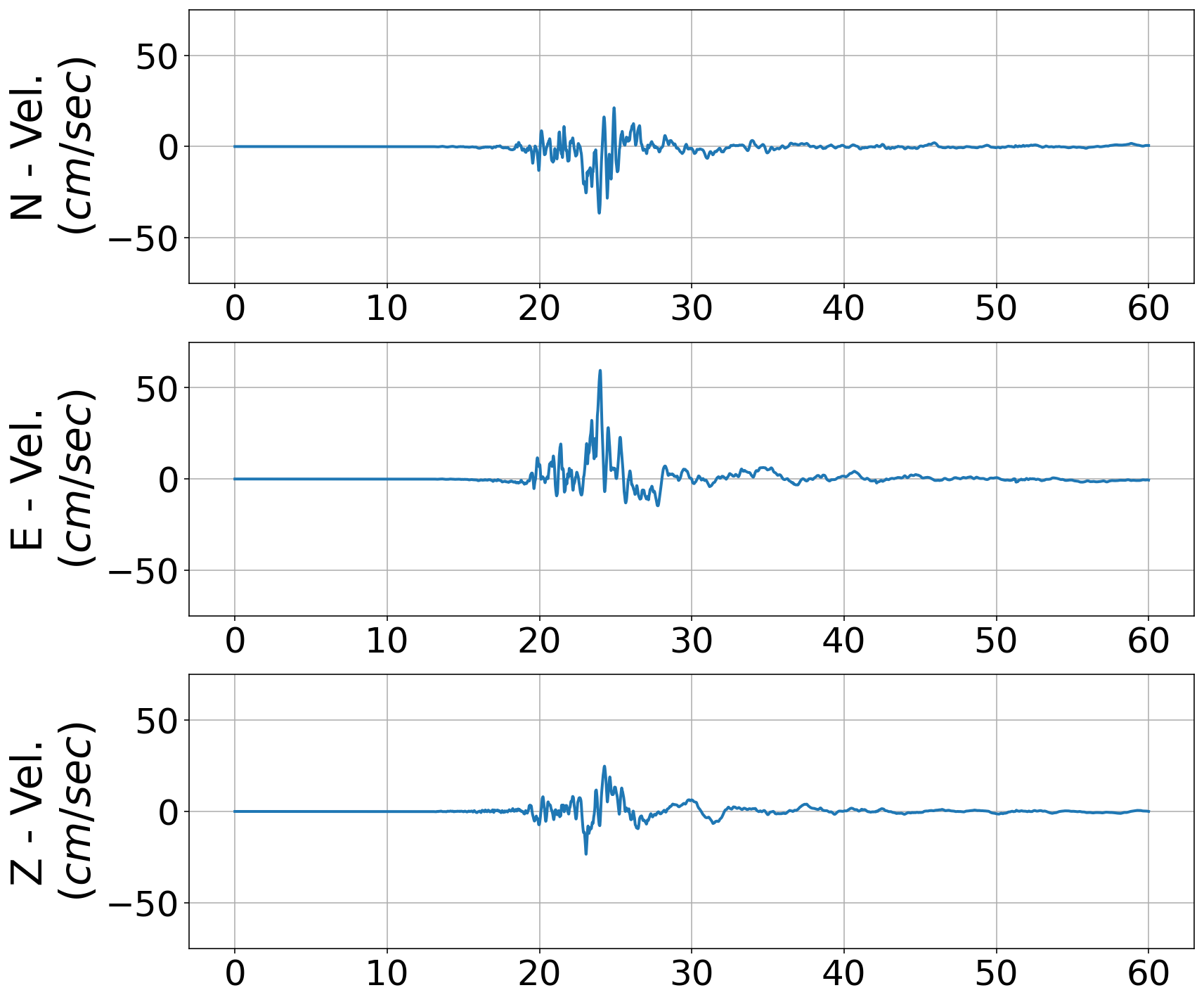}
        \end{subfigure}
        \begin{subfigure}{\linewidth}
          \centering
          \caption{}
          \includegraphics[width=0.9\linewidth]{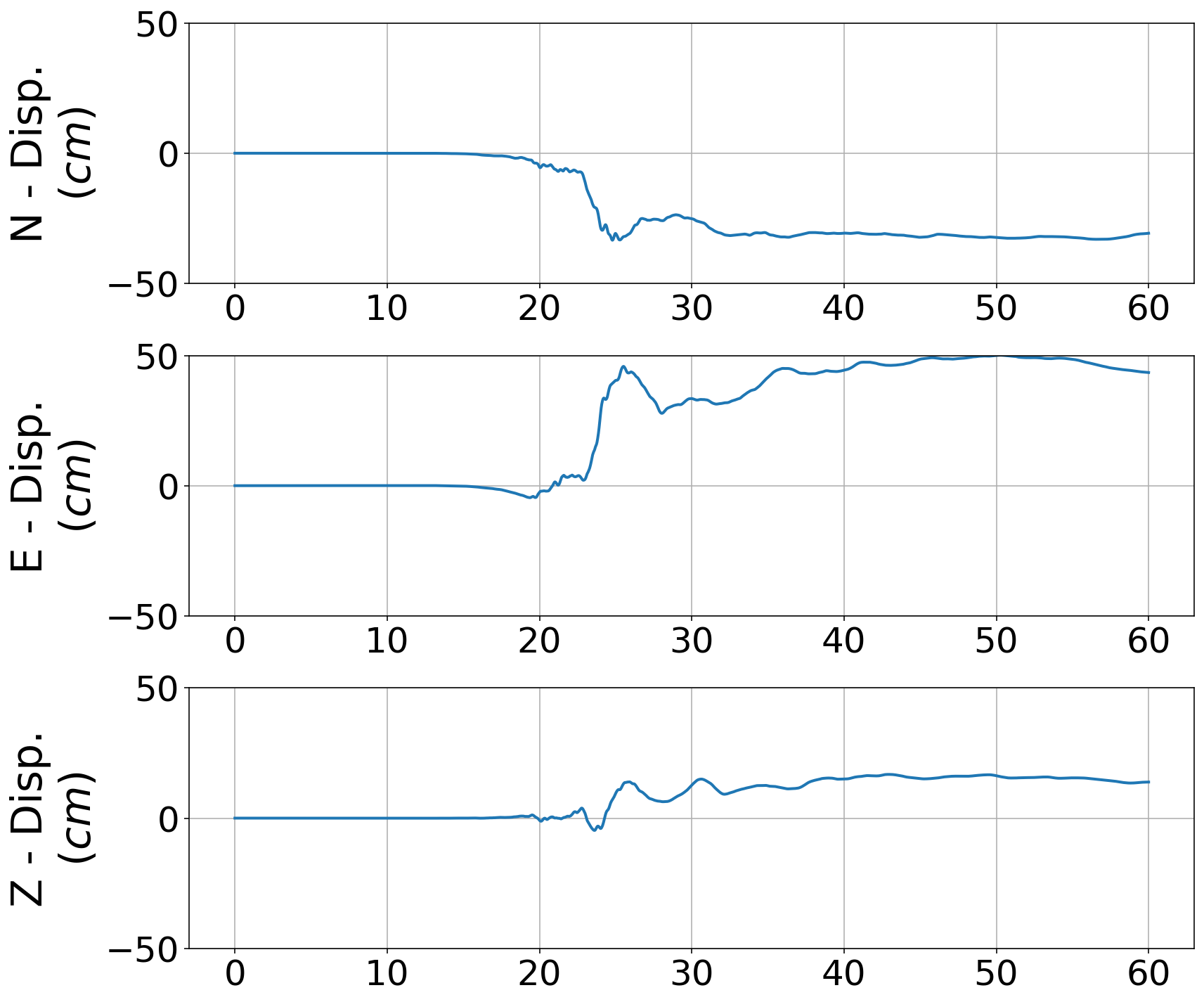}
        \end{subfigure}
        \caption{Baseline correction of ground motions recorded at station HWA075 during the \mbox{M\,6.9} Chishang earthquake, using timing selections based only on the north–south component. (a) acceleration time histories, (b) velocity time histories, and (c) displacement time histories.}  
        \label{fig:examp_timing_wrong}
    \end{minipage}
\end{figure}

To illustrate this, Figure~\ref{fig:examp_timing_correct} presents the baseline-corrected time histories from station HWA075 for the \mbox{M\,6.9} Chishang earthquake, using time selections based on all three components. In contrast, Figure~\ref{fig:examp_timing_wrong} shows the results when timing was selected using only the North–South (N–S) component.

As shown in Figure~\ref{fig:examp_timing_wrong}, relying solely on the N–S component can lead to inaccuracies. In this case, the start of strong shaking, and consequently $t_1$, was overestimated, while the vertical component clearly indicates an earlier onset. More importantly, performing baseline correction independently on each component prevents cross-validation of the selected time instances. Reviewing the East–West (E–W) and vertical (Z) components reveals residual trends after $t_2$, suggesting that the time window was not optimally chosen. In contrast, Figure~\ref{fig:examp_timing_correct} shows that coordinating the $t_1$ and $t_2$ timing across all components results in corrected displacements that are flat and stable after the strong motion phase.

Comparison with observations from the nearby GA34 mobile GNSS station, 1.13 km away from the strong motion sensor, further validates the proposed approach. The GNSS station recorded a permanent displacement of 21.5~cm in the South direction. The proposed multi-component method yielded an estimate of 19.9~cm, whereas the single-component method produced a significantly overestimated displacement of 30~cm.

\section{Ground Motion Characteristics} \label{sec:gm_characteristics}

Using the proposed approach, time histories from both the \mbox{M\,6.5} Guanshan foreshock and the \mbox{M\,6.9} Chihshang mainshock were processed, and the relevant intensity measures were computed in the time domain to avoid artifacts in the frequency domain caused by fling-step discontinuities.

\subsection{Peak-to-Peak Velocity Distribution}

In this study, the horizontal peak-to-peak velocity ($PPV_H$) is defined as the maximum Euclidean norm of the difference between horizontal velocity vectors at two distinct time instances, $t'$ and $t''$, separated by less than 2.5 seconds:
\begin{equation} \label{eq:ppv_h}
    PPV_H = \max\left( \left\| \mathbf{v}(t') - \mathbf{v}(t'') \right\|_2 \right), \quad \text{for} \quad 0 < t' < t'', \quad t' - t'' < 2.5,
\end{equation}
where $\mathbf{v}(t) = [v_{N\text{-}S}(t), v_{E\text{-}W}(t)]$ is the horizontal velocity vector at time $t$. The use of the vector norm ensures that $PPV_H$ characterizes the same velocity pulse across both components. The 2.5-second window is used to limit the maximum pulse period to approximately 5.0 seconds.  A sensitivity analysis confirmed that the selected time window had a negligible impact on the computed $PPV_H$. 
Based on this formulation, the component-wise peak-to-peak velocity ($PPV_i$) is computed as the absolute velocity difference over the same time interval used for $PPV_H$:
\begin{equation}
    PPV_i = | v_i(t') - v_i(t'') |, \quad \text{where} \quad t', t'' \text{ are from Equation~\ref{eq:ppv_h}},
\end{equation}
with $i \in \{N\text{-}S, E\text{-}W\}$.

\begin{figure}[htbp!]
    \centering
    \includegraphics[width=\textwidth]{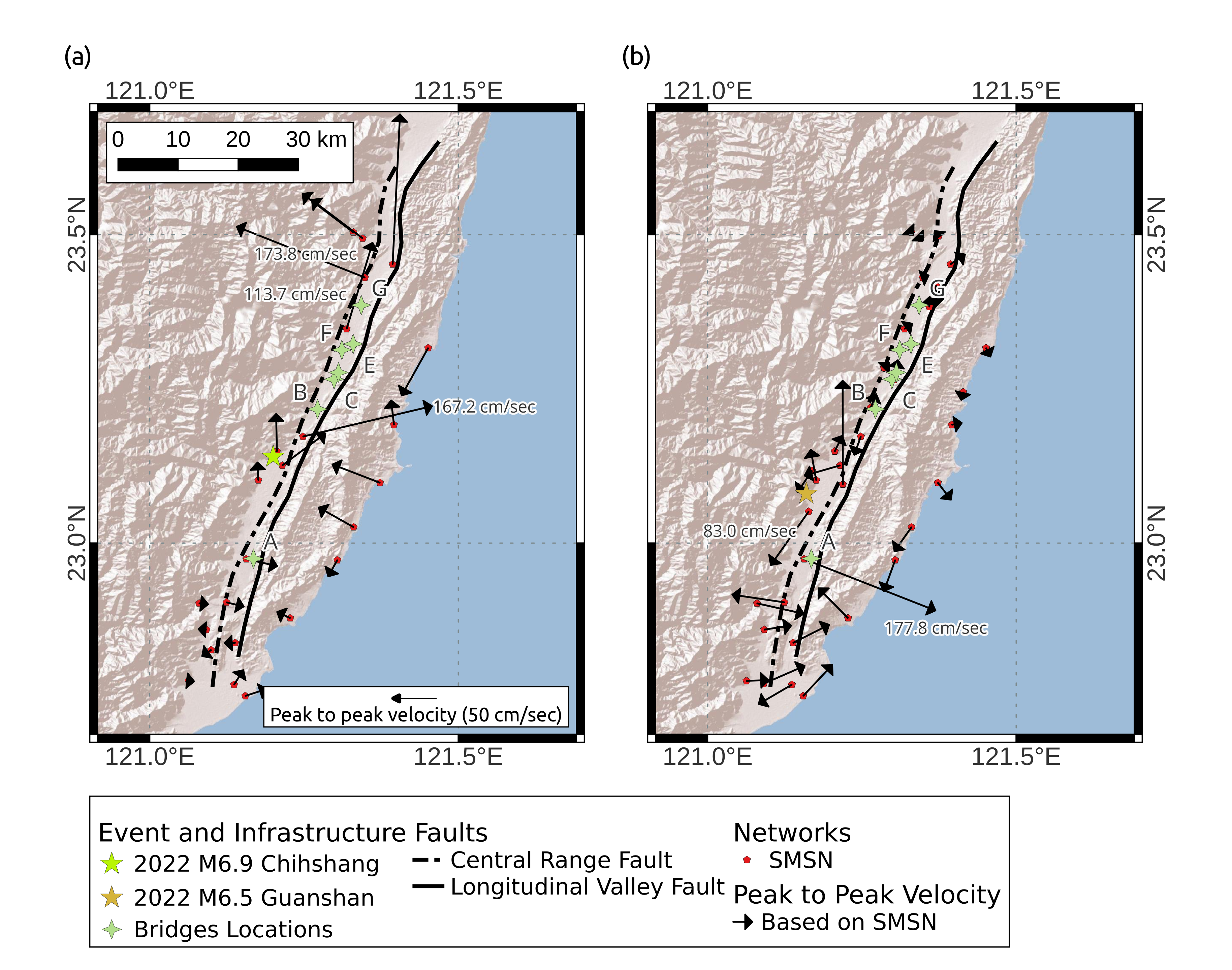}
    \caption{Peak-to-peak velocity spatial distribution for the 2022 Chisang earthquake sequence. (a)   \mbox{M\,6.5} Guanshan foreshock and \mbox{M\,6.9} Chishang mainshock}
    \label{fig:ppv_map}
\end{figure}

Figure~\ref{fig:ppv_map} presents the spatial distribution and orientation of $PPV_H$ vectors for both the \mbox{M\,6.5} Guanshan foreshock (Figure~\ref{fig:ppv_map}a) and the \mbox{M\,6.9} Chihshang mainshock (Figure~\ref{fig:ppv_map}b). Despite the difference in magnitude, both events generated high peak-to-peak velocities, with the maximum $PPV_H$ in each event exceeding 170~cm/s.

In the \mbox{M\,6.5} Guanshan event, the largest $PPV_H$ values were concentrated in the southwestern side of the rupture, while in the \mbox{M\,6.9} Chihshang event, the highest $PPV_H$ values were observed along the northeastern side. 
In both cases, the largest peak-to-peak velocities occurred away from the respective epicenters and were aligned with the fault-normal direction, which suggests that they were the result of forward directivity. Similar findings were identified in the recent study by \cite{Poulos2024} investigating the directionality of SDOFs' response for the same earthquake sequence.

\subsection{Tectonic Offset Distribution}

The permanent displacement ($D_p$) is defined as the mean value of the displacement time history over the final one second of each component:
\begin{equation}
    D_p = \text{mean}\left( d(t) \right) \quad \text{for} \quad t \in [t_e - 1, t_e],
\end{equation}
where $d(t)$ is the displacement time history and $t_e$ is the total duration of the record.

To support comparison and aid in validating the proposed baseline correction method, permanent displacement measurements from high-rate GNSS stations operated by \cite{CWA2012} were also included in the analysis.

\begin{figure}[htbp!]
    \centering
    \includegraphics[width=\textwidth]{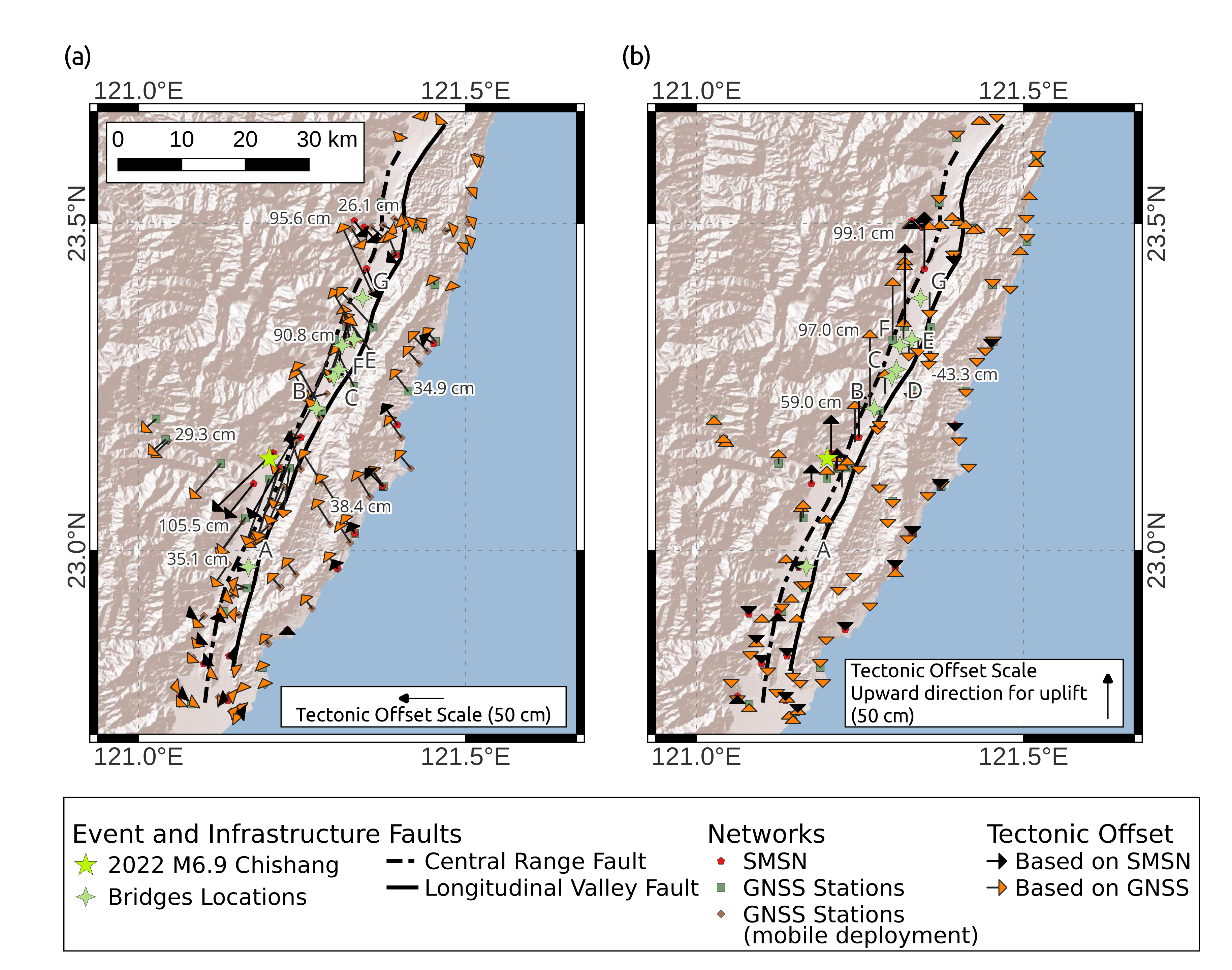}
    \caption{Horizontal tectonic offset for the 2022 Chisang earthquake sequence. (a) \mbox{M\,6.5} Guanshan foreshock and (b)  \mbox{M\,6.9} Chishang mainshock}
    \label{fig:tect_slip_horiz_map}
\end{figure}

\begin{figure}[htbp!]
    \centering
    \includegraphics[width=\textwidth]{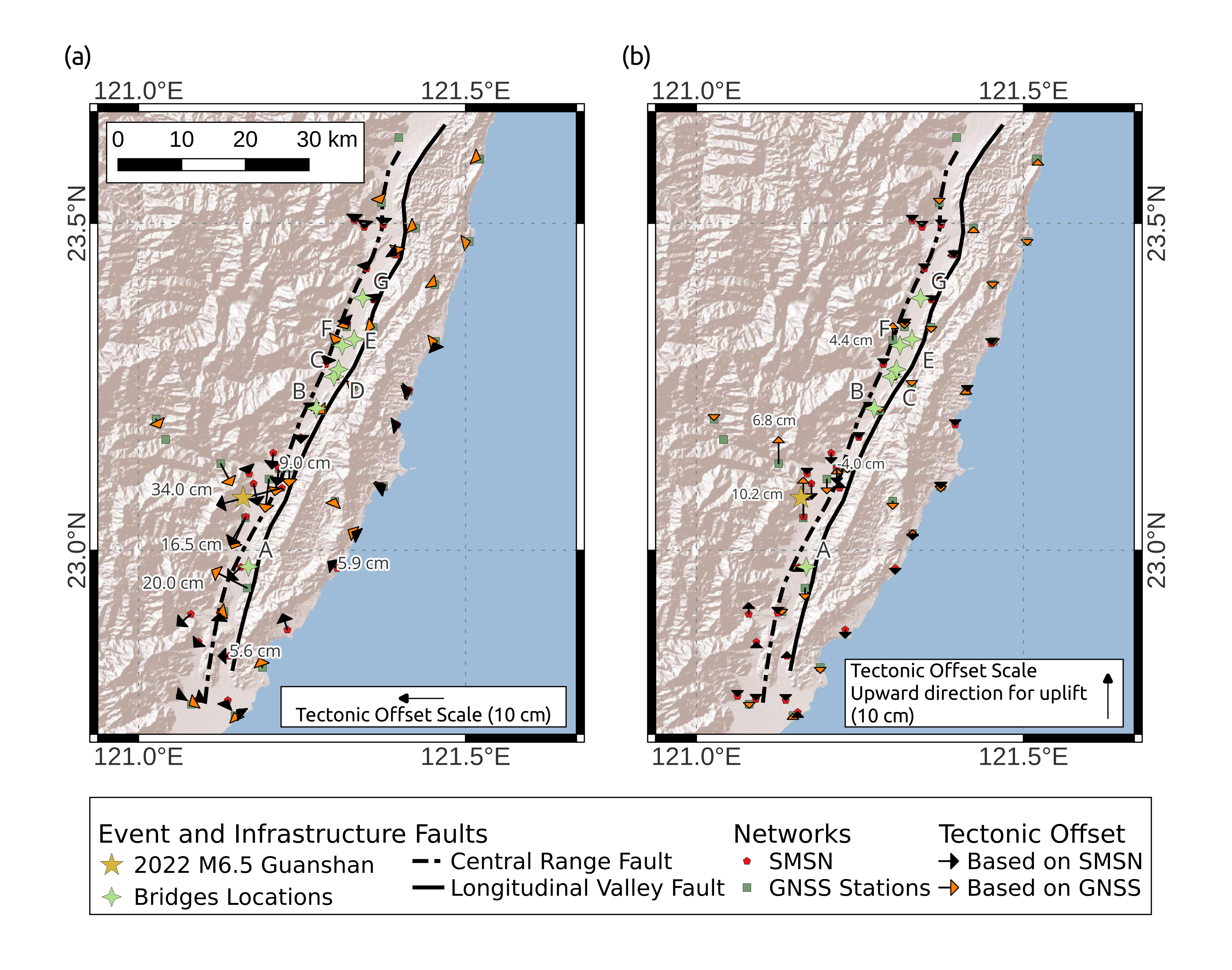}
    \caption{Vertical tectonic offset for the 2022 Chisang earthquake sequence. (a) \mbox{M\,6.5} Guanshan foreshock and (b) \mbox{M\,6.9} Chishang mainshock}
    \label{fig:tect_slip_vert_map}
\end{figure}

Figures~\ref{fig:tect_slip_horiz_map}a and \ref{fig:tect_slip_horiz_map}b show the horizontal components of tectonic offset for the \mbox{M\,6.5} Guanshan and \mbox{M\,6.9} Chihshang earthquakes, respectively. Consequently, Figures~\ref{fig:tect_slip_vert_map}a and \ref{fig:tect_slip_vert_map}b depict the vertical components of the tectonic offset for the same two events.

Regarding the tectonic offset, the \mbox{M\,6.5} Guanshan produced relatively smaller displacements with the largest tectonic offset located close to the epicenter, resulting in approximately $10 cm$ of uplift of the Longitudinal Valley (west of CRF) and  $34 cm$ of left-lateral oblique shortening. 

The \mbox{M\,6.9} Chishang mainshock produced significantly larger displacements in both horizontal and vertical components compared to the foreshock. In the horizontal direction, two prominent features are observed: a pronounced left-lateral offset of the Longitudinal Valley Fault and a shortening component across the Coastal Range (east of CRF).  

The largest left-lateral offset, on the order of 100 cm, is observed near the epicenter close to the township of Chishang. A similar offset is evident further east in the Central Range (the mountainous region further east from CRF), although with a reduced amplitude of approximately 30 cm.  The shortening component is most pronounced on the northeastern side of the rupture near the township of Yuli, where horizontal displacements reach up to 95 cm. This shortening effect persists until the coast, with amplitudes decreasing to around 35 cm near the shoreline.  

In the vertical component, significant uplift ranging from 60 to 100 cm is observed along the Longitudinal Valley northeast of the rupture. In contrast, the Coastal Range exhibits moderate subsidence, ranging from 20 to 40 cm. Both horizontal and vertical displacements are notably smaller on the southwestern side of the rupture.  

For both events, the tectonic offsets inferred from strong motion recordings exhibit consistent patterns with GNSS displacements and the kinematic inversion results from \cite{Tang2023}. This consistency highlights the capability of strong motion data, when properly baseline-corrected, to capture tectonic displacements in both horizontal and vertical components.

\subsubsection{Validation with GNSS}

\begin{figure}[htbp!]
    \centering
    \begin{subfigure}[t]{0.32\textwidth}
        \caption{}
        \includegraphics[height = 0.85\textwidth]{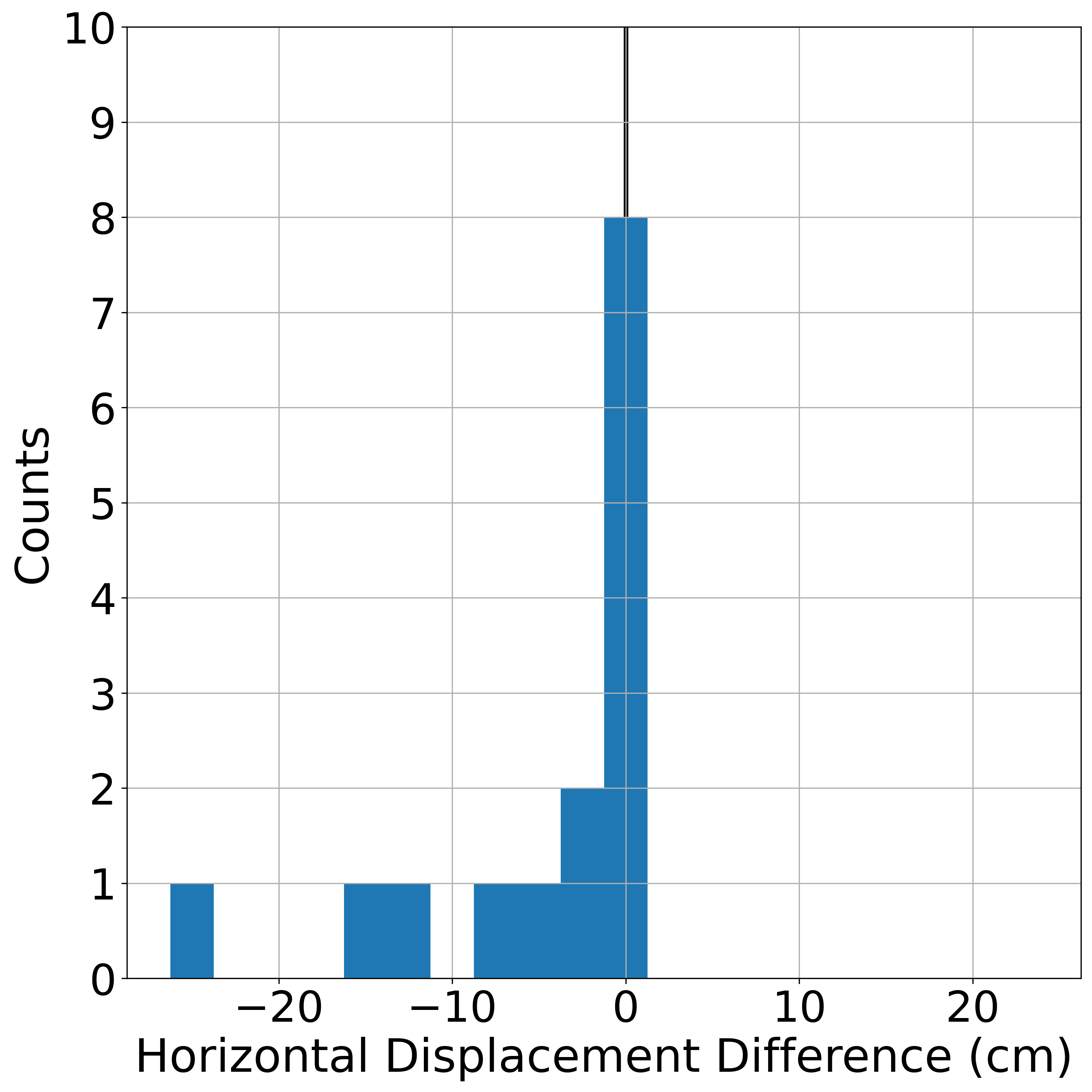}
    \end{subfigure} 
    \begin{subfigure}[t]{0.32\textwidth}
        \caption{}
        \includegraphics[height = 0.85\textwidth]{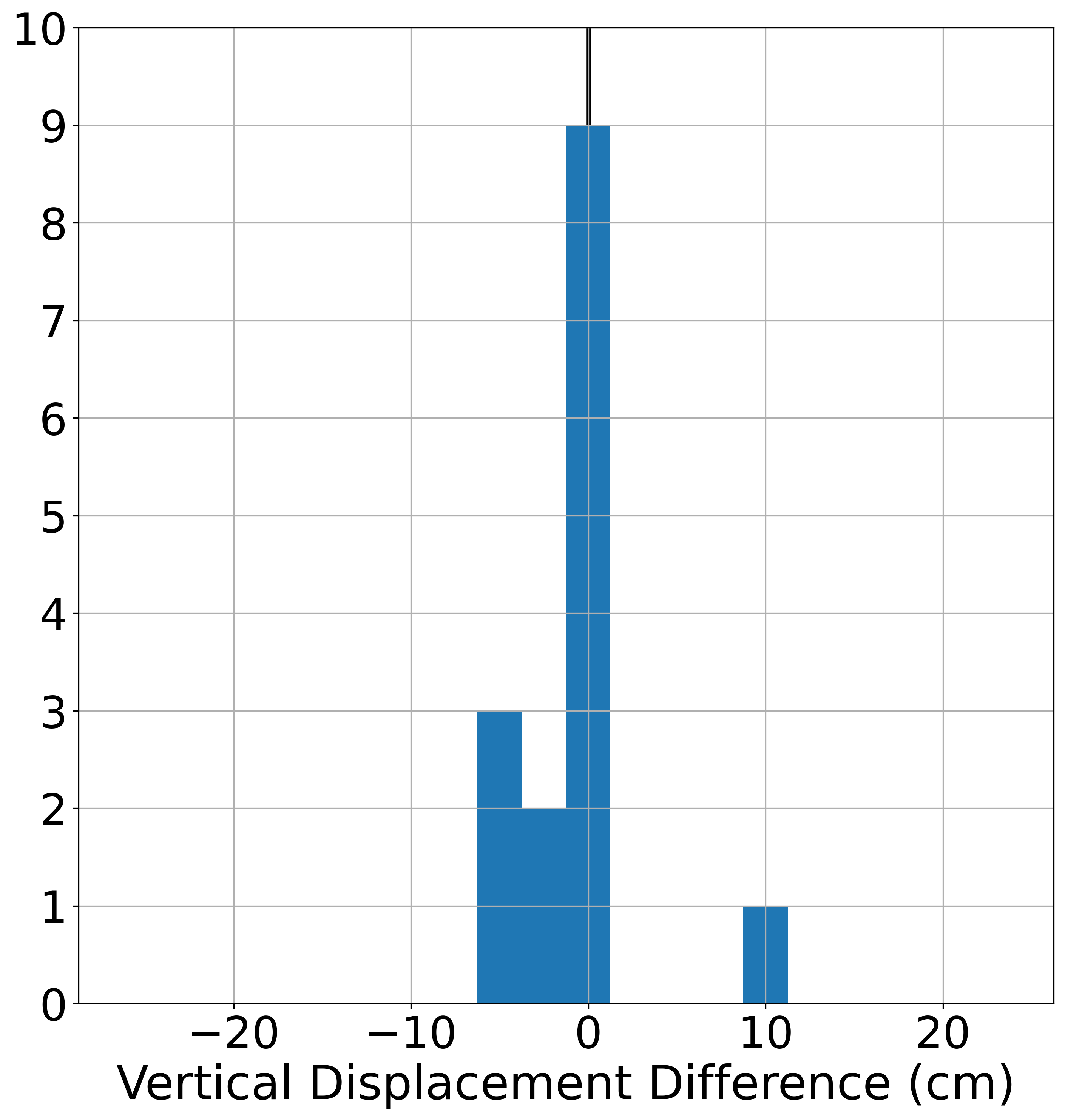}
    \end{subfigure}  
    \begin{subfigure}[t]{0.32\textwidth}
        \caption{}
        \includegraphics[height = 0.85\textwidth]{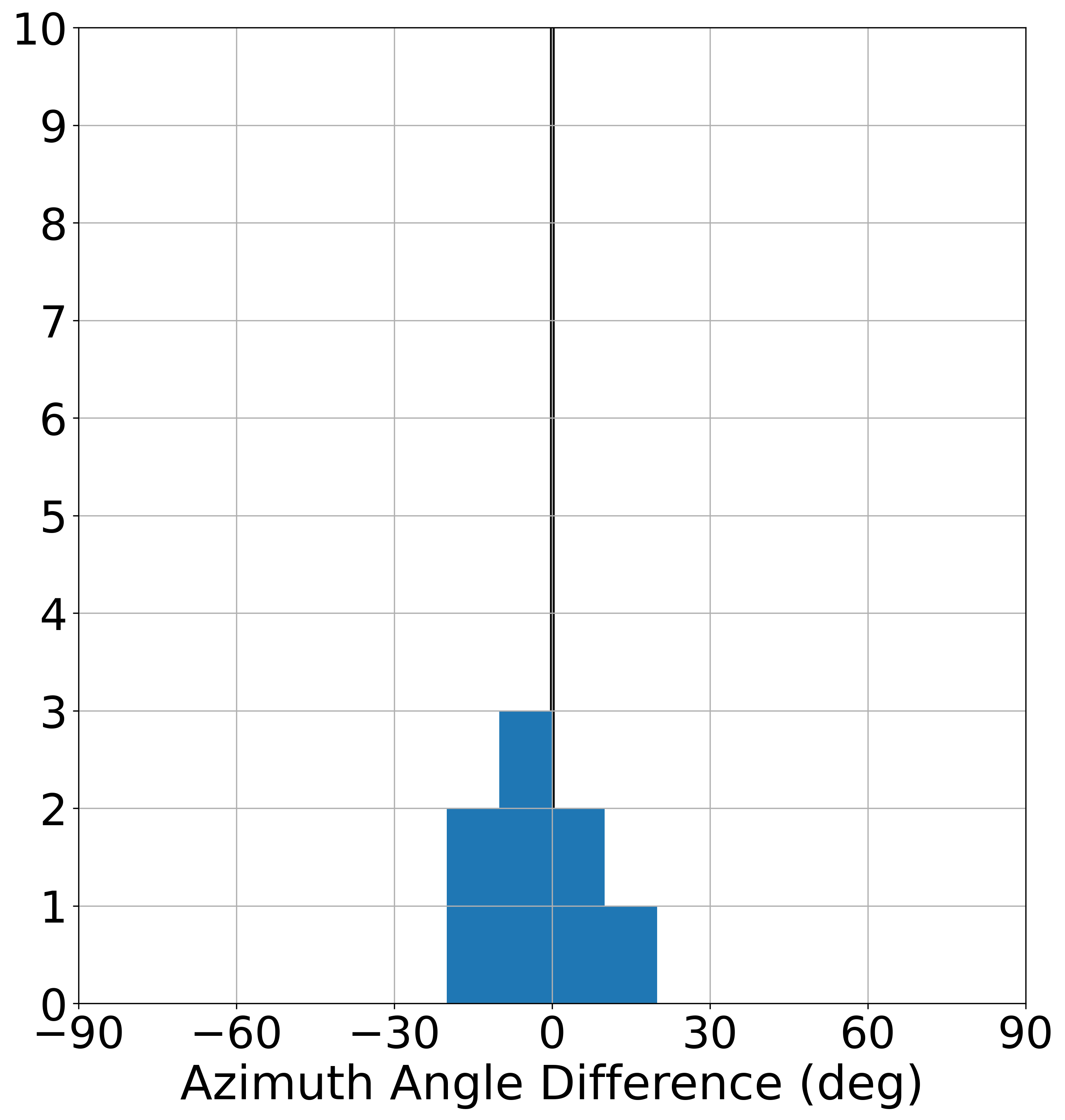}
    \end{subfigure}  
    \caption{Comparison of tectonic offsets estimated from the Strong Motion Seismic Network (SMSN) using the proposed baseline correction method against nearby Global Navigation Satellite System (GNSS) stations. Shown are differences in (a) horizontal offset, (b) vertical offset, and (c) azimuthal orientation of the displacement vector. In all cases, differences are computed as the GNSS estimate minus the SMSN estimate.}
    \label{fig:valid_offset}
\end{figure}

A more quantitative evaluation of the estimated tectonic offsets from the processed SMSN records was conducted by comparing the horizontal and vertical displacement amplitudes, as well as the azimuthal orientation of the displacement vectors, between nearby SMSN and GNSS stations (Figure \ref{fig:valid_offset}). The comparison was performed for both events using station pairs located within 1~km of each other and positioned on the same side of the fault, based on pre-event fault trace mapping by the Central Geological Survey of Taiwan. The second criterion was imposed to avoid comparisons across the fault trace, where coseismic offsets may differ significantly. In all cases, the displacement difference is defined as the GNSS estimate minus the SMSN estimate

Figure~\ref{fig:valid_offset} shows that the SMSN-derived displacements are generally consistent with those obtained from GNSS, though a few outliers are present.

For the horizontal component, most SMSN estimates fall within 5~cm of the corresponding GNSS values, with only three cases exceeding 10~cm in absolute difference. The largest discrepancy occurs between SMSN station TTN020 and GNSS station TTCS, both located within 2.5 km of the epicenter. At this location, the horizontal displacement estimated from SMSN data is 74 cm, while the GNSS-derived displacement is 94 cm, representing an approximate 20\% underestimation by the SMSN record. The other two significant deviations occur near the coast: between SMSN-TTN057 and GNSS-DULI, and between SMSN-TTN002 and GNSS-S220. In both cases, the GNSS stations recorded displacements of approximately 30 cm, whereas the SMSN-derived estimates were closer to 15 cm. Notably, in all three instances, the azimuthal orientations of the displacement vectors between SMSN and GNSS remained consistent.

The comparison of vertical displacements shows even closer agreement, with all but one station pair exhibiting differences of less than 5~cm. The largest discrepancy occurs between SMSN-HWA073 and GNSS-E801, where the SMSN-derived uplift is 99 cm compared to 88 cm from the GNSS data. However, significant variation in vertical displacement is also observed between adjacent GNSS stations suggesting that this discrepancy may be partially attributed to the spatial variability in vertical tectonic deformation, rather than inaccuracies in the baseline correction procedure.

Finally, the comparison of azimuthal orientation shows that the majority of SMSN-derived directions fall within $\pm$10\textdegree of the GNSS measurements, with a few cases showing differences of up to 20\textdegree. This suggests that the proposed method is generally successful in capturing the direction of permanent tectonic displacement.

Overall, the comparison indicates that the proposed baseline correction approach is capable of recovering displacement time histories that include tectonic deformation, albeit with some discrepancies when compared to more robust methods such as GNSS. While relying solely on this method to infer absolute tectonic displacements may not be ideal, it nonetheless offers a valuable tool, particularly in providing strong-motion acceleration time histories that reasonably capture the fling step, which can be used to evaluate structural response in near-fault regions.

\subsection{Velocity Pulse Classification}

\begin{figure}[htbp!]
    \centering
    \begin{subfigure}[t]{0.32\textwidth}
        \caption{}
        \includegraphics[height = 0.85\textwidth]{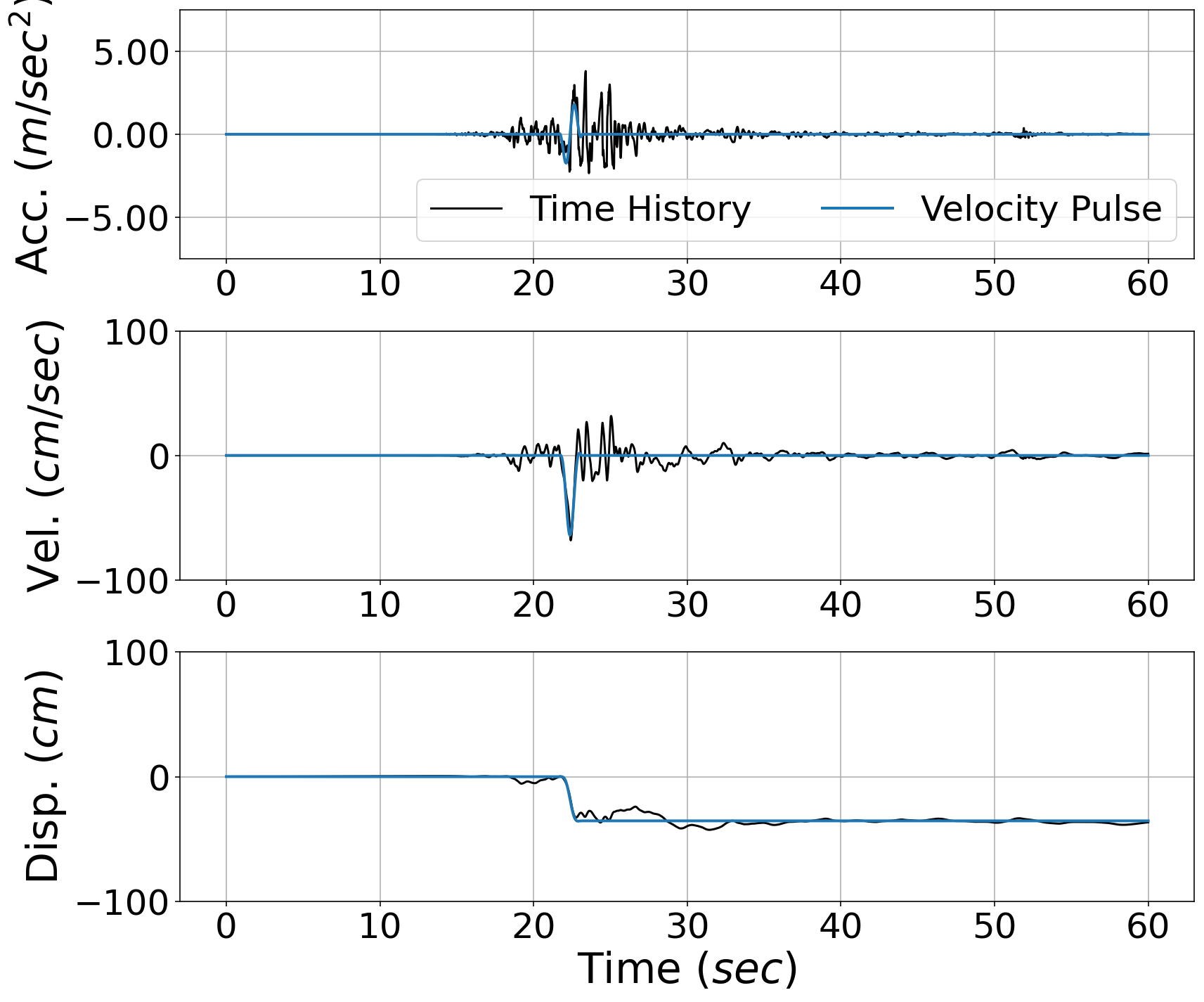}
    \end{subfigure} 
    \begin{subfigure}[t]{0.32\textwidth}
        \caption{}
        \includegraphics[height = 0.85\textwidth]{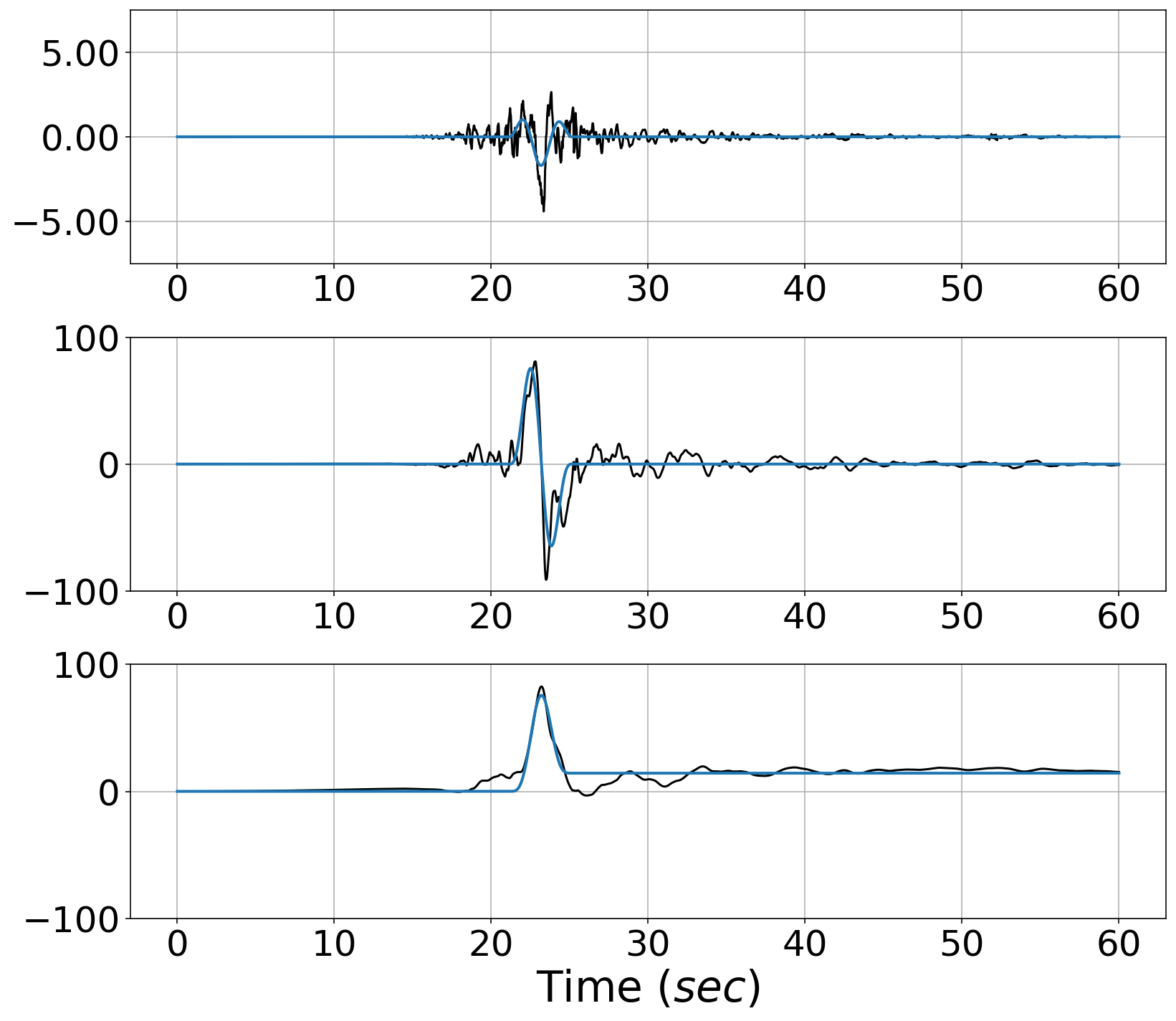}
    \end{subfigure}  
    \begin{subfigure}[t]{0.32\textwidth}
        \caption{}
        \includegraphics[height = 0.85\textwidth]{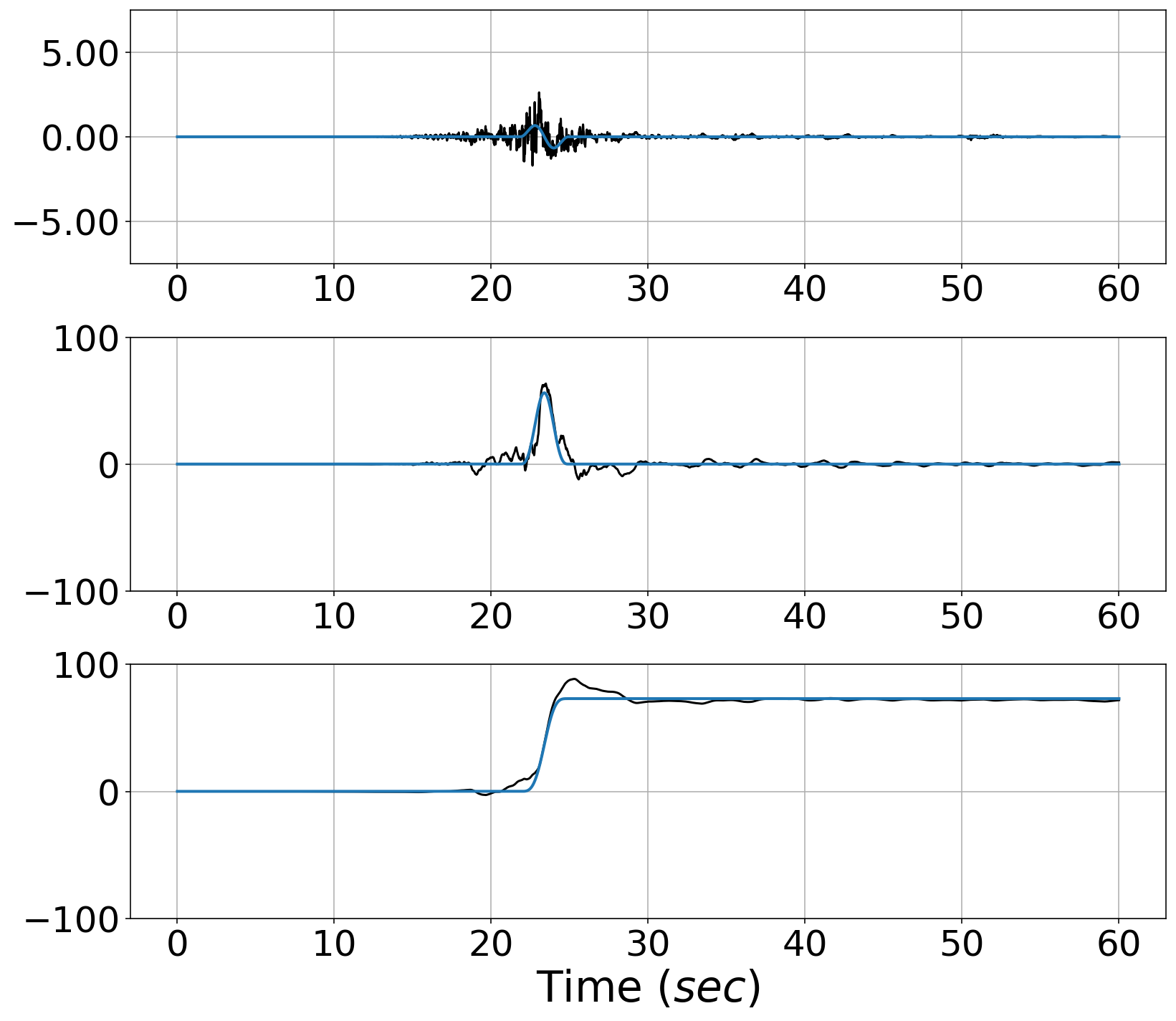}
    \end{subfigure}  
    \caption{Velocity pulse fit applied to the three-component time history of the \mbox{M\,6.9} Chishang earthquake recorded at the HWA054 station ($R_{\mathrm{rup}} = 2.3$~km), 
    (a) North-South, (b) East-West, and (c) Vertical component.}
    \label{fig:vel_pulse_fit}
\end{figure}

The idealized representation of velocity pulses proposed by \citet{Mavroeidis2003}, given by:
\begin{equation}
    v_p(t) = \frac{A}{2}\left(1 + \cos\left(2 \pi \frac{f_p}{\gamma} (t - t_p) \right)\right) \cos\left(2 \pi f_p (t - t_p) + \nu \right) \quad \text{for} \quad t \in \left[t_p - \frac{\gamma}{2 f_p}, t_p + \frac{\gamma}{2 f_p} \right]
\end{equation}
is used in this study to classify the processed ground motion records based on their velocity pulse characteristics. In this formulation, $A$ denotes the amplitude, $f_p$ is the pulse frequency, $t_p$ is the time at which the pulse is centered, $\gamma$ controls the effective duration (or number of cycles) of the pulse, and $\nu$ represents a phase shift.
Different from the original formulation,  $\gamma$ was allowed to take values smaller than one, enabling the model to better represent single-sided velocity pulses.
Following the pulse fitting, the residual displacement $D_{\mathrm{res}}$ is defined as the integral of the velocity pulse from the start to the end of the record: ($D_{res} = \int_0^{+\infty} v_p(t) dt$).
Figure~\ref{fig:vel_pulse_fit} shows a representative fit of the idealized velocity pulse model to the three-component velocity time histories recorded at station HWA054 during the \mbox{M\,6.9} Chihshang earthquake. 
This example shows that the implemented formulation is capable of both capturing velocity pulses introduced by fling effects visible in the North-South and Vertical components, as well as velocity pulses introduced by forward directivity visible in the East-West component.


Using the extracted velocity-pulse parameters ($A$, $\gamma$, $D_{\mathrm{res}}$) together with the peak ground displacement ($\text{PGD}$) and peak ground velocity ($\text{PGV}$) of each processed record, each time history was classified as containing: a fling pulse, dynamic pulse, or neither, based on the following decision tree:

\forestset{
  my node/.style={
    draw,
    rounded corners,
    align=center,
    inner sep=3pt,
    font=\small
  }
}

\begin{figure}[htbp]
\centering
\begin{forest}
  for tree={my node, l sep=10pt, s sep=12pt}
  [{$\gamma < 2.0$ ?}
    [Not classified, edge label={node[midway,left]{No}}]
    [{$\displaystyle\frac{D_{\mathrm{res}}}{\text{PGD}} > 0.5$ ?},
      edge label={node[midway,right]{Yes}}
      [{$\displaystyle\frac{A}{\text{PGV}} > 0.5$ ?},
        edge label={node[midway,left]{No}}
        [Not classified, edge label={node[midway,left]{No}}]
        [\textbf{Dynamic pulse}, edge label={node[midway,right]{Yes}}]
      ]
      [\textbf{Fling-step pulse}, edge label={node[midway,right]{Yes}}]
    ]
  ]
\end{forest}
\end{figure}

The term dynamic pulse is used to describe large, single-cycle velocity pulses resulting in small residual displacement. This terminology is preferred over the more specific term forward-directivity pulse to encompass pulses observed in both horizontal and vertical components. In the horizontal direction, most dynamic pulses are attributed to forward rupture directivity, as they typically appear on the fault-normal component in the direction of rupture propagation (Figure~\ref{fig:ppv_map}). However, vertical pulses may arise from other rupture-related mechanisms.

The rationale behind the classification criteria is as follows. The condition $\gamma < 2$ is used to exclude records exhibiting multi-cycle pulses. The ratio of residual to peak displacement, $D_{\mathrm{res}}/\text{PGD}$, is used to identify records with significant permanent ground displacement. Values greater than 0.5 are interpreted as indicative of fling-step pulses.  For records where $D_{\mathrm{res}}/\text{PGD} \le 0.5$, the ratio of velocity pulse amplitude to peak ground velocity ($A/\text{PGV}$) is examined. If the velocity pulse amplitude is comparable to the overall PGV (i.e., $A/\text{PGV} > 0.5$), the record is classified as containing dynamic pulses.

The thresholds used for $D_{\mathrm{res}}/\text{PGD}$ and $A/\text{PGV}$ were selected empirically based on visual agreement between the classification results and trends observed in the processed time histories.



Using the classification scheme described above, 27 records were identified as containing fling-step pulses, 19 in the horizontal component and 8 in the vertical component, while 31 records were classified as exhibiting dynamic pulses, 19 in the horizontal and 12 in the vertical direction. An important question relevant for structural design that has been underexplored is the timing of these velocity pulses.
To address this, we define the quarter-pulse arrival ($t'_p = t_p - \gamma/(4 f_p)$) and compute the normalized-cumulative-squared velocity to define the timing of the velocity pulse relative to the rest of the strong ground motion:
\begin{equation}
    i_v(t'_p) = \frac{\int_0^{t'_p} v^2(t) dt}{\int_0^{+\infty} v^2(t) dt}
\end{equation}
The quarter-pulse arrival and the associated normalized-cumulative-squared velocity are chosen over the pulse start ($t_p - \gamma/(2 f_p)$) or center time ($t_p$) because the former may occur near the beginning of the record, making difficult to define the associated $i_v$, while the $i_v$ associated with the latter can be skewed due to the presence of the pulse itself.

\begin{figure}[htbp!]
    \centering
    \includegraphics[width=0.4\textwidth]{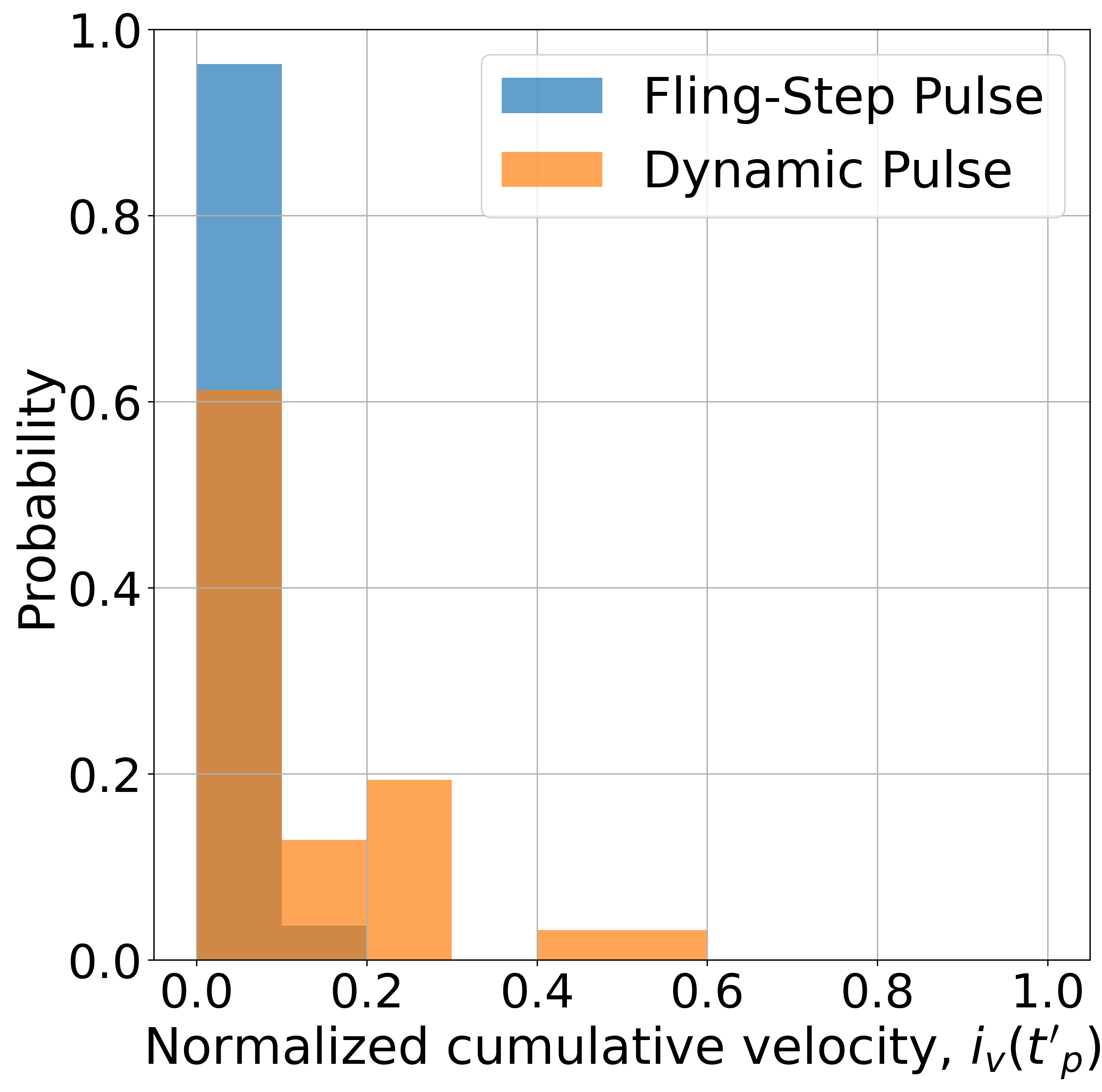}
    \caption{Timing of fling-step and dynamic velocity pulses, characterized using the normalized cumulative squared velocity ($i_v$) at the quarter-duration time of the velocity pulse ($t_p'$).}
    \label{fig:vel_pulse_time}
\end{figure}

Figure~\ref{fig:vel_pulse_time} compares the relative timing of fling-step and dynamic pulses in terms of $i_v(t_p')$. The results show that both types of pulses typically occur early in the strong shaking. Fling-step pulses are tightly clustered in the $i_v = 0.0$--$0.1$ interval, while dynamic pulses span a broader range of $i_v = 0.0$--$0.3$. The early arrival of fling pulses is consistent with theoretical expectations, as the static offset component is associated with the near-field term of the displacement field due to a double-couple source, which propagates at the P-wave velocity \citep{Lay1995}.

\begin{figure}[htbp!]
    \centering
    \includegraphics[width=0.4\textwidth]{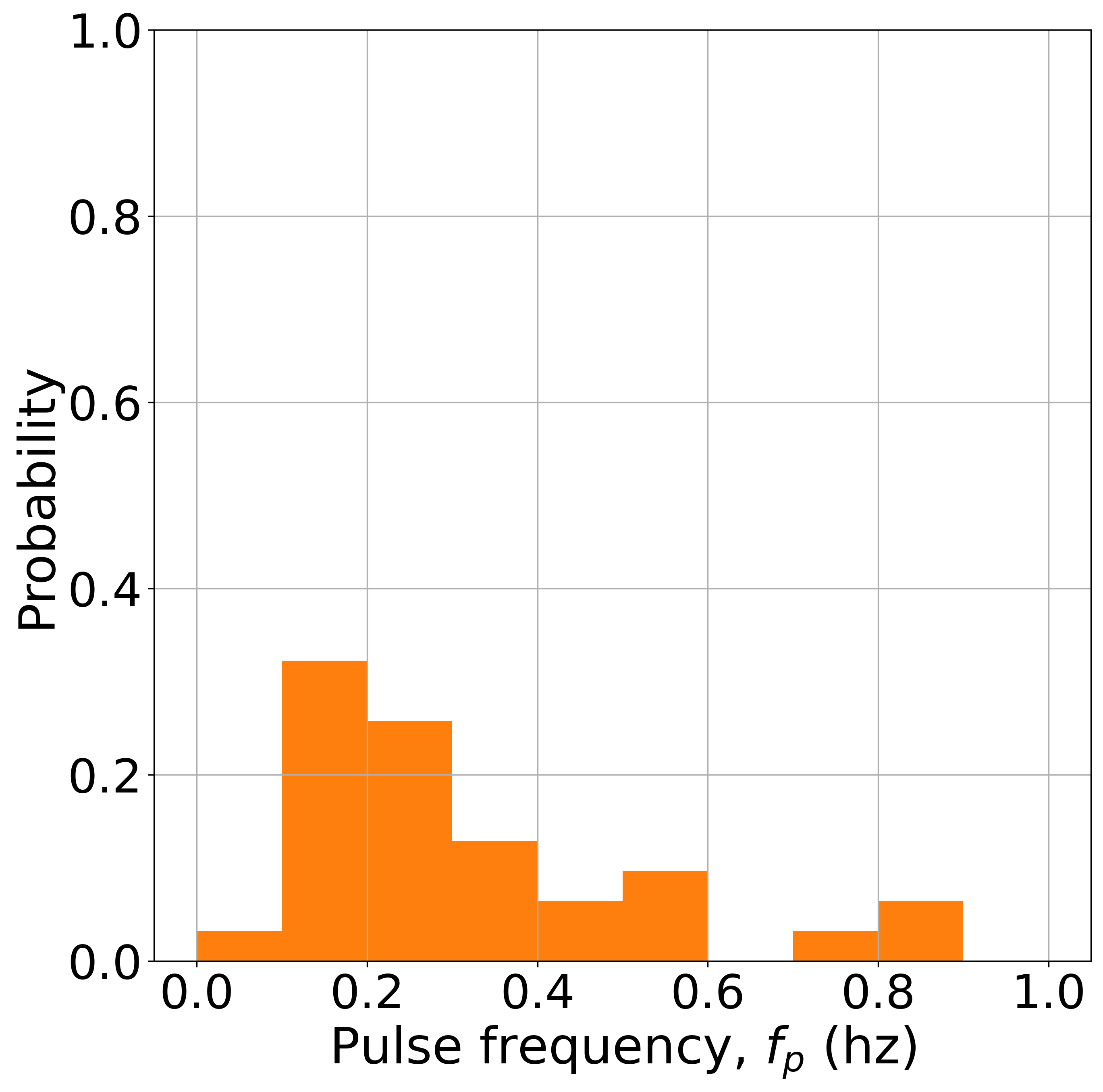}
    \caption{Distribution of dominant pulse frequencies, $f_p$ for dynamic velocity pulses.}
    \label{fig:vel_pulse_period}
\end{figure}

Figure~\ref{fig:vel_pulse_period} shows the distribution of dominant frequencies associated with dynamic velocity pulses from both events. Fling-step pulses are excluded from this analysis, as they primarily consist of single-sided pulses for which a dominant frequency is not well defined.
The dynamic pulses span a broad frequency range, with most falling between 0.1 and 0.6~Hz. Both horizontal and vertical components are included in the comparison, as they exhibit similar frequency characteristics. The average dominant frequency is approximately 0.32~Hz, which is consistent with pulse period estimates reported by \citet{Shahi2014} for earthquakes of similar magnitude.

\section{Correlation with Structural Damage} \label{sec:structural_damage}

The GEER team surveyed seven bridges~\citep{Asimaki2024}, and the damage observed ranged from superficial cracking and minor joint repairs to the complete collapse of two multi-span bridges. Five bridges suffered mild to moderate damage that required joint repair, superficial element reconstruction, or span realignment. When the GEER team was deployed in October 2022, the damaged bridges were in various states of repair. Although the M6.5 and M6.9 earthquakes contributed to the observed damage, the M6.9 event was responsible for the most severe damage and collapses.

The seismic performance of the seven bridges provides critical data for performance-based design, which informs both the seismic resilience of the existing bridge stock and the development of functional recovery models. Beyond conventional metrics, the observed seismic bridge performance also offers insight into near-fault ground motion effects, because all seven bridges are located within ten kilometers of the fault. Accordingly, in this section we classify the observed bridge damage mechanisms and relate them to the intensity of near-fault ground motions using the PPV computation approach described above. We also relate the bridge damage mechanisms to traditional ground motion intensity parameters [i.e., peak ground acceleration (PGA) and peak ground velocity (PGV)].

The bridge locations are provided in Figures~\ref{fig:ppv_map},~\ref{fig:tect_slip_horiz_map},~and~\ref{fig:tect_slip_vert_map}, which show strong motion station PPV and tectonic offsets. Table~\ref{tab:bridgedamage} summarizes the bridge coordinates, distance to the nearest strong motion station, bridge azimuth, and recorded intensity measures (PGA, PGV, and PPV) on the bridges. The resolved bridge components were computed using the abutment coordinates to determine the normal and parallel directions relative to the bridge azimuth. The normal component corresponds to the short axis of the bridge (perpendicular to traffic), while the parallel component aligns with the long axis (direction of traffic).

\begin{table}[]
    \centering
    \caption{Caption}
    \begin{tabular}{c c | c c | c c | c c}
    \hline
    \thead{Label} & \thead{Name} & \thead{PPV$_\mathrm{BN}$\\(cm/s)} & \thead{PPV$_\mathrm{BP}$\\(cm/s)} & \thead{PGA$_\mathrm{BP}$\\(g)} & \thead{PGA$_\mathrm{BN}$\\(g)} & \thead{PGV$_\mathrm{BP}$\\(cm/s)} & \thead{PGV$_\mathrm{BN}$\\(cm/s)}\\
    \hline
     A & Baohua  & 7.2	& 39.7	& 0.14	& 0.09	& 27.4	& 13.0\\
     B & Luntain  &  -109.2	& 126.7 &	0.33 &	0.55 & 	84.5 &	71.9\\
    C & Zhangfu  & -110.2	& -28.1 & 	0.53 &	0.40 &	51.7 &	98.1\\
     D & Dongli RR  & -12.9	& 112.9 &	0.43 &	0.52 &	98.8 &	54.4\\
     E & Yuli (2X)   &  -113.2	& 10.3 &	0.49 &	0.49 &	56.7 &	96.5\\
     F & Kecheng RR   & 40.5	& 106.2 &	0.54 &	0.42 &	87.2 &	63.6\\
      G & Gaoliao  & -112.6	& 15.8 &	0.48 & 	0.50 &	56.9 &	95.4\\
    \hline
    \end{tabular}
    \label{tab:bridgeintensity}
\end{table}

The GEER team observed limited evidence of liquefaction or lateral spreading at the surveyed sites \citep{Carey2023}, consistent with observations made by other teams prior to the arrival of the GEER team. Therefore, the observed bridge damage was primarily associated with structural kinematics rather than ground failure. Local site effects, such as amplification of ground motion or degradation of soil strength and stiffness, may have contributed to performance, but these effects were not easily observable in the field.

\subsection{PPV Comparisons and Observed Damage}

\begin{figure}[htbp!]
    \centering
      \includegraphics[width=0.85\textwidth]{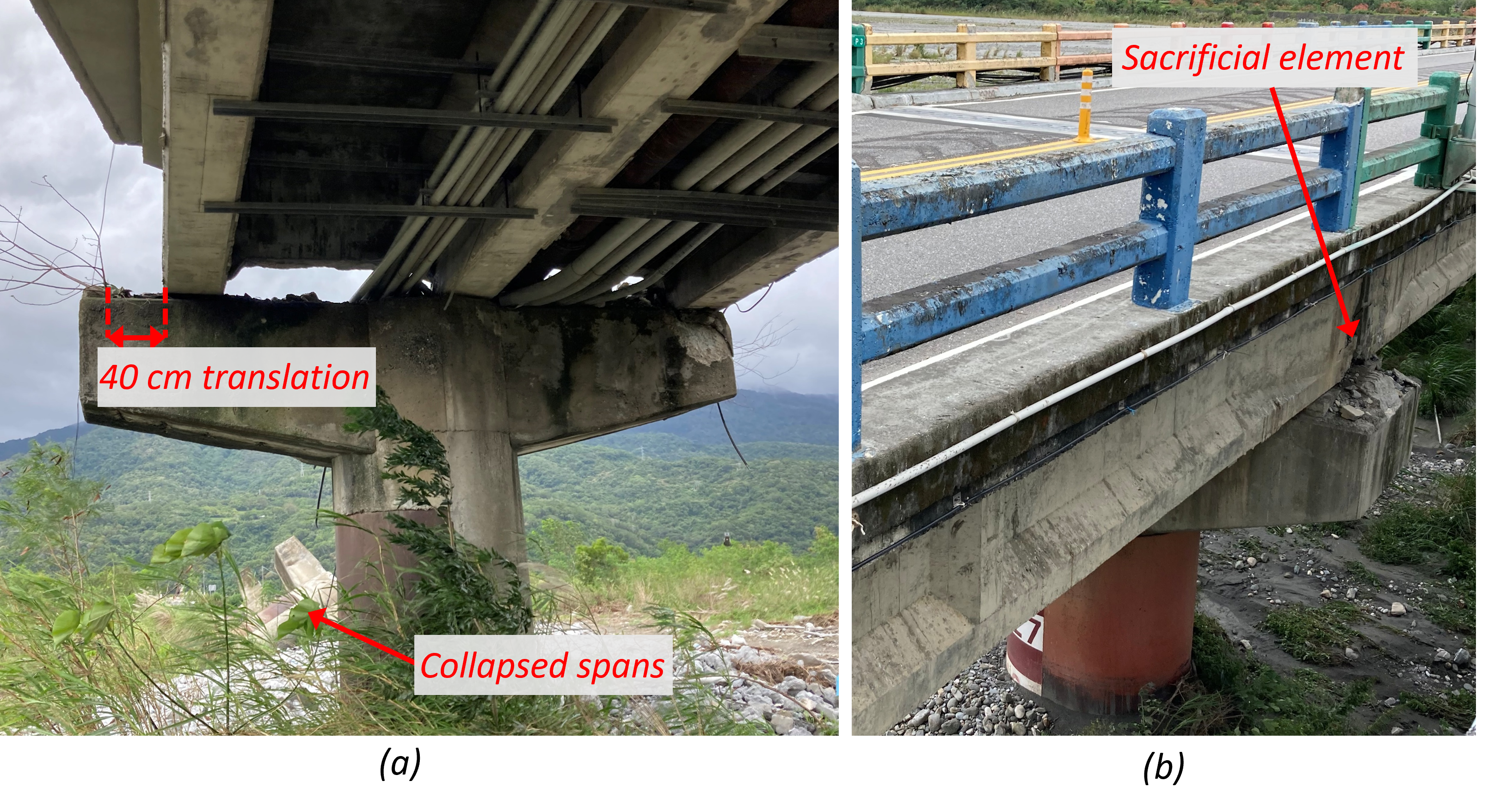}
    \caption{(a) Translation of reinforced concrete deck girders and collapsed spans at the Lutain bridge and (b) damage to lateral restraining sacrificial elements at the twin Yuli bridges.}
    \label{fig:DamangeToSelectBridges}
\end{figure}

The largest normal-direction PPV (PPV$_{BN}$) demands were recorded at the Lutain, Zhangfu, Yuli, and Gaoliao bridges. Among these, the Lutain and Gaoliao bridges experienced the most severe damage, including multi-span failures that resulted in full closure. The likely failure mechanism involved unseating of reinforced concrete deck girders, which lacked lateral restraints or shear keys to resist transverse displacements~--- a type of girder translation consistent with near-fault, pulse-like ground motion effects \citep{fenves1998behavior,liao2004comparison,Loh2002}. Figure~\ref{fig:DamangeToSelectBridges}a shows the Lutain Bridge deck displaced approximately 40 cm off its elastomeric bearing pads, with complete unseating of adjacent spans leading to collapse.

The Zhangfu and twin Yuli bridges also experienced high PPV$_{BN}$ demands, but their girders remained seated due to sacrificial elements that limited lateral displacement (see Figure~\ref{fig:DamangeToSelectBridges}b). The existence of sacrificial elements probably reflects differences in construction eras or seismic retrofit strategies, particularly at the twin Yuli bridges, which cross the Eurasian–Philippine Sea Plate suture zone.

In contrast, the Dongli and Kecheng railroad bridges were oriented nearly parallel to the Longitudinal Valley Fault. Consequently, the highest PPV demands acted along the longitudinal bridge axis. The Dongli Bridge suffered significant joint and abutment damage, with several spans displaced out of alignment. Field evidence suggests that the failures were caused by longitudinal demands approaching 112 cm/s. When the GEER team arrived, bridge repairs were already underway. The Kecheng Bridge, although subjected to similar longitudinal PPV demands, sustained only minor damage, probably due to a greater joint capacity, which could be incorporated as a seismic mitigation measure given its proximity to the boundary of the Eurasia-Philippine plate.

Finally, the Baohua Road Bridge, located south of the epicenter, shared a similar orientation to other roadway bridges but did not experience strong directivity effects. The damage to the bridge was minor and it remained operational after the earthquakes. Here, PPV$_{BN}$ was only 7 cm/s, compared to values exceeding 110 cm/s at the most heavily damaged bridges. Although the longitudinal PPV$_{BP}$ demand reached 40 cm / s, retrofits were implemented to reduce longitudinal displacements before the earthquakes, which probably contributed to the good seismic performance of the bridge.

\subsection{Comparison with Traditional Intensity Measures}

Traditional engineering approaches typically quantify intensity measures (e.g., PGA, PGV) for each horizontal component separately, and then combine them through rotated directions (RotD50, RotD100) or geometric means in the case of evolutionary measures. Although this approach produces maximum envelopes, it obscures the temporal and directional characteristics of near-fault motions. In particular, pulse-like directivity effects can result in strongly asymmetric demands across bridge axes, which are not captured when combining components.
Table~\ref{tab:bridgedamage} presents PGA and PGV values for the resolved bridge-normal and bridge-parallel components. Although the largest PGA and PGV values still aligned with the same bridge directions as the PPV analysis, the magnitudes were more evenly distributed across both axes. This makes it more difficult to identify the primary loading direction responsible for damage. For example, at the Gaoliao Bridge, PGV in the normal direction was 95 cm/s, while in the parallel direction it was 57 cm/s, a relatively modest difference compared to the PPV-based values of 113 cm/s and 14 cm/s, respectively.

The approach described herein provides a more physically consistent representation of bridge demands because it accounts for the fact that peak intensities are unlikely to occur simultaneously in both principal directions. By resolving intensities into normal and parallel components, this method better reflects the actual loading conditions that bridges experienced during the earthquakes.

\begin{figure}[htbp!]
    \centering
      \includegraphics[width=0.85\textwidth]{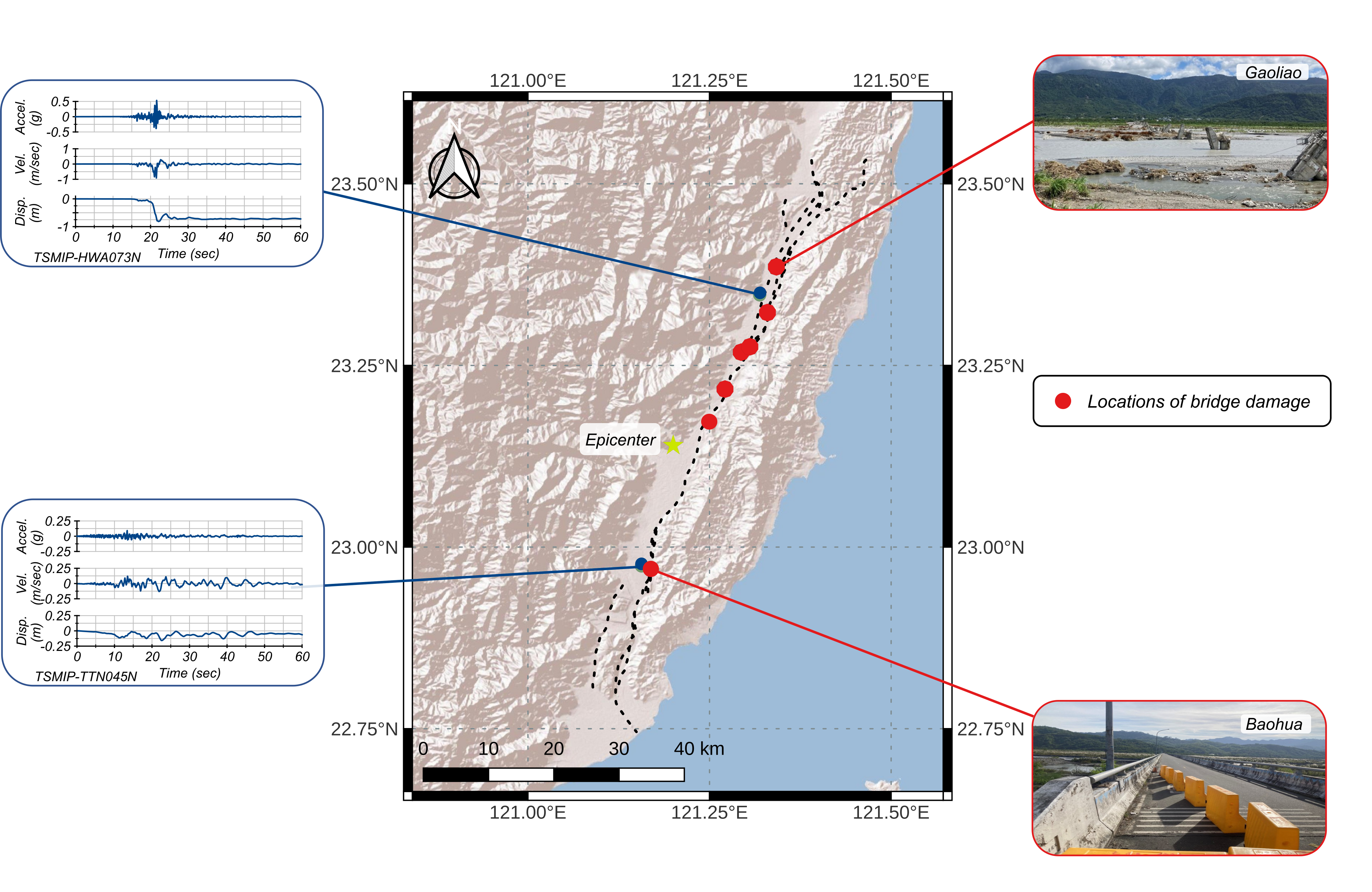}
    \caption{Location of bridge damage, select ground motion recordings, and epicenter of the $M 6.9$ Chishang earthquake.}
    \label{fig:BridgeDamage}
\end{figure}

Major damage occurred to bridge systems during the M6.9 Chishang mainshock, including the collapse of two multi-span bridges north of the epicenter. In Figure~\ref{fig:BridgeDamage} are the locations of the damaged bridges and select ground motion recordings for the M6.9 Chishang mainshock. The bridges that experienced the most severe damage, including total failure, had longitudinal axes perpendicular to the Central Range Fault and the direction of rupture propagation. Bridge failure or translation of deck girders is consistent with known near-fault effects, particularly those involving pulse-like ground motions \citep{fenves1998behavior,liao2004comparison,Loh2002}. The orientation of the bridges in the Hualien–Taitung Valley made them particularly vulnerable to the effects of dynamic and fling velocity pulses.

At the affected bridge sites, there was no evidence of ground failure of liquefaction, lateral spreading, or downdrag-related foundation damage. Instead, failure mechanisms are likely linked to unseating of reinforced concrete deck girders, which lacked lateral restraints or shear keys to resist transverse motion. The unseating likely caused plastic hinge formation in substructure elements or global rotational instability due to insufficient lateral capacity. At several sites, deck girders were visibly translated longitudinally and transversally off their elastomeric bearing pads, with measured displacements of at least 40 cm. Bridges with axes parallel to the rupture direction showed different damage patterns, including abutment pounding, deck misalignment, and buckling. 

The Gaoliao Bridge was located near the 95 cm shortening zone (Figure \ref{fig:tect_slip_horiz_map}b), and is the northernmost bridge from the epicenter (Figure~\ref{fig:BridgeDamage}) that experienced damage. Constructed in 1991, this 879-meter-long bridge experienced widespread collapse, with 21 of its 22 spans failing and plastic hinges forming in 20 or 21 piers \citep{Asimaki2024}. Post-event surveys and photographic evidence indicate that the bridge lacked lateral restraint for the deck girders, which were seated on elastomeric bearing pads~--- an observation supporting the conclusion that near-fault effects led to girder unseating and structural collapse.

In contrast, the Baohua Bridge, located south of the epicenter (Figure~\ref{fig:BridgeDamage}), experienced much less damage. This 680-meter-long structure, older than the Gaoliao Bridge and similarly lacking lateral restraints, showed only minor damage to pipe connections and deck guardrails. The absence of girder unseating at this site likely reflects differences in the intensity or orientation of the dynamic pulses, consistent with a location outside the primary rupture direction.

\section{Discussion and Conclusions}

This study extends the baseline correction methods proposed by \citet{Boore2001EffectEarthquake, Boore2002CommentsEarthquake} by applying synchronous baseline correction to all three components of strong-motion accelerograms to more accurately preserve near-fault ground motion characteristics, particularly permanent tectonic offsets and velocity pulses associated with rupture directivity and the fling effect.
The synchronous selection of time intervals for correcting spurious acceleration offsets is based on the assumption that instrument tilting is the primary source of such artifacts. By applying the same correction windows across all three components, the method provides an internal consistency check, allowing for the identification of cases in which single-component corrections would lead to erroneous estimates of permanent displacement.

The 2022 Chihshang earthquake sequence, including the \mbox{M\,6.5} Guanshan foreshock and the \mbox{M\,6.9} Chihshang mainshock, provided a valuable dataset due to the dense distribution of near-fault strong-motion stations. Analysis of the processed records revealed key near-field features, including pronounced velocity pulses and significant permanent displacements.
Large peak-to-peak velocity (PPV) amplitudes were observed during both events, primarily in the fault-normal component. The foreshock produced the largest PPV values on the southwestern segment of the fault, whereas the mainshock generated large PPV values along the northeastern portion. The Guanshan foreshock was associated with small to moderate tectonic displacements concentrated near the epicentral region, while the Chihshang mainshock produced substantial horizontal and vertical displacements that extended northeastward along the fault trace.

Close alignment between the tectonic offsets inferred from strong-motion records and those measured by GNSS further validates the proposed approach. The comparison shows that the baseline correction method yields reliable estimates of permanent displacement, typically within 5~cm in both the horizontal and vertical directions, and accurately captures the azimuthal orientation of the displacement vectors observed by GNSS. While some discrepancies persist, the method provides a valuable tool for recovering strong-motion time histories that retain near-fault effects, making them suitable for use in seismic design.

A classification algorithm is proposed to categorize the processed ground motions based on the presence of near-fault characteristics. Each time history is classified as exhibiting either a dynamic velocity pulse (typically associated with rupture directivity), a fling-step pulse, or neither. 
Analysis of the extracted velocity pulse features indicates that fling-step pulses tend to occur early in the ground shaking, whereas dynamic pulses appear over a broader temporal window. Furthermore, the dominant frequencies of the dynamic pulses are shown to be consistent with previously reported empirical relationships for pulse-like ground motions.

Bridge damage observed after the earthquake sequence strongly correlates with the distribution of high peak-to-peak velocities and large tectonic offsets recorded during the \mbox{M\,6.9} Chihshang mainshock. The majority of damaged bridges were oriented approximately normal to the fault, making them particularly susceptible to fling-step effects.
These observations underscore the critical role of near-fault ground motion features in influencing structural performance and highlight the need for processing methods that accurately preserve such effects in time histories used for engineering analysis and design.

\section*{Acknowledgements}

The authors gratefully acknowledge the support of the Geotechnical Extreme Events Reconnaissance (GEER) Association. Financial support for GEER is provided by the National Science Foundation (NSF) through the Geotechnical Engineering Program under Grant No. CMMI-1266418. 
Any opinions, findings, conclusions, or recommendations expressed in this material are those of the authors and do not necessarily reflect the views of the National Science Foundation. We would like to thank our other GEER team members, Adda Athanasopoulos-Zekkos, Fernando Estefan Garcia, Brian Gray, and Chukwuebuka Nweke, who helped us with the post-earthquake reconnaissance work as well as the in-country support provided by members of the National Center for Research on Earthquake Engineering (NCREE) and Academia Sinica. We would particularly like to thank Kuo-Fong Ma, of Academia Sinica, who helped us access and understand the recorded ground motions.

\section*{Declaration of Competing Interests}

The authors declare that there are no conflicts of interest to report.

\section*{Data and Resources}

The signal processing code, supporting scripts for figure generation, and processed time histories are available at:
\url{https://github.com/glavrentiadis/geer_2022Chishang_ground_motions}

The processed time histories are publicly accessible through DesignSafe at:
\url{https://www.designsafe-ci.org/data/browser/public/designsafe.storage.published/PRJ-5934}

\bibliography{references_GL, references_other}

\begin{thebibliography}{31}
\providecommand{\natexlab}[1]{#1}
\providecommand{\url}[1]{\texttt{#1}}
\expandafter\ifx\csname urlstyle\endcsname\relax
  \providecommand{\doi}[1]{doi: #1}\else
  \providecommand{\doi}{doi: \begingroup \urlstyle{rm}\Url}\fi

\bibitem[Abrahamson et~al.(1999)Abrahamson, Bardet, Boulanger, Bray, Chan, Chang, Chen, Harder, Huang, Huang, Ju, Kramer, Kuo, Lee, Lin, Loh, Ou, Seed, Shen, Sitar, Stewart, Teng, Sun, and Wright]{Abrahamson1999GEER002}
Norman Abrahamson, Jean-Pierre Bardet, Ross Boulanger, Jonathan Bray, Yu-Wen Chan, Chien-Yu Chang, Chia-Hung Chen, Lee Harder, An-Bin Huang, Sheng Huang, Jen-Wen Ju, Steven Kramer, Min-Ju Kuo, Wei-Fang Lee, Hsing-Lung Lin, Chin-Hsiung Loh, Ching-Yu Ou, Raymond Seed, Ji-Der Shen, Nicholas Sitar, Jonathan Stewart, Li~Teng, Jiing-I Sun, and Robert Wright.
\newblock Preliminary geotechnical earthquake engineering observations of the september 21, 1999, ji-ji, taiwan earthquake: A report of the taiwan-us geotechnical earthquake engineering reconnaissance team.
\newblock Technical Report GEER-002, GEER Association, October 1999.

\bibitem[Ancheta et~al.(2014)Ancheta, Darragh, Stewart, Seyhan, Silva, Chiou, Wooddell, Graves, Kottke, Boore, Kishida, and Donahue]{Ancheta2014}
Timothy~D. Ancheta, Robert~B. Darragh, Jonathan~P. Stewart, Emel Seyhan, Walter~J. Silva, Brian~S.J. Chiou, Katie~E. Wooddell, Robert~W. Graves, Albert~R. Kottke, David~M. Boore, Tadahiro Kishida, and Jennifer~L. Donahue.
\newblock {NGA-West2 database}.
\newblock \emph{Earthquake Spectra}, 30\penalty0 (3):\penalty0 989--1005, 2014.
\newblock ISSN 87552930.
\newblock \doi{10.1193/070913EQS197M}.

\bibitem[Asimaki et~al.(2024)Asimaki, Mason, Athanasopoulos-Zekkos, Carey, Garcia, Gray, Lavrentiadis, Nweke, Chou, Wu, Lee, Chang, Chen, Lin, Lin, Lin, Wang, Lin, and Lee]{Asimaki2024}
Domniki Asimaki, H.~Benjamin Mason, Adda Athanasopoulos-Zekkos, Trevor Carey, Fernando~E. Garcia, Brian Gray, Grigorios Lavrentiadis, Chukwuebuka~C. Nweke, Chung-Che Chou, Chiun-Lin Wu, Zheng-Kuan Lee, Wei-Kuang Chang, Kuan-Yu Chen, Shih-Jung Lin, Chi-Hao Lin, Che-Min Lin, Kuo-lung Wang, Jun-Tin Lin, and Yi-Hsuan Lee.
\newblock Geotechnical engineering reconnaissance of the 2022 chihshang earthquake sequence.
\newblock Technical report, GEER, 2024.

\bibitem[Boore(2001)]{Boore2001EffectEarthquake}
D.~M. Boore.
\newblock {Effect of Baseline Corrections on Displacements and Response Spectra for Several Recordings of the 1999 Chi-Chi, Taiwan, Earthquake}.
\newblock \emph{Bulletin of the Seismological Society of America}, 91\penalty0 (5):\penalty0 1199--1211, 10 2001.
\newblock ISSN 0037-1106.
\newblock \doi{10.1785/0120000703}.
\newblock URL \url{https://pubs.geoscienceworld.org/bssa/article/91/5/1199-1211/102918}.

\bibitem[Boore(2002)]{Boore2002CommentsEarthquake}
D.~M. Boore.
\newblock {Comments on Baseline Correction of Digital Strong-Motion Data: Examples from the 1999 Hector Mine, California, Earthquake}.
\newblock \emph{Bulletin of the Seismological Society of America}, 92\penalty0 (4):\penalty0 1543--1560, 5 2002.
\newblock ISSN 0037-1106.
\newblock \doi{10.1785/0120000926}.
\newblock URL \url{https://pubs.geoscienceworld.org/bssa/article/92/4/1543-1560/120799}.

\bibitem[Carey et~al.(2023)Carey, Mason, Asimaki, Athanasopoulos-Zekkos, Garcia, Gray, Lavrentiadis, and Nweke]{Carey2023}
Trevor~J Carey, H~Benjamin Mason, Domniki Asimaki, Adda Athanasopoulos-Zekkos, Fernando~E Garcia, Brian Gray, Grigorios Lavrentiadis, and Chukwuebuka~C Nweke.
\newblock The 2022 chihshang, taiwan, earthquake: Initial geer team observations.
\newblock \emph{Journal of Geotechnical and Geoenvironmental Engineering}, 149\penalty0 (5):\penalty0 02823002, 2023.

\bibitem[{Central Weather Administration (CWA, Taiwan)}(2012)]{CWA2012}
{Central Weather Administration (CWA, Taiwan)}.
\newblock Central weather administration seismographic network.
\newblock \url{https://gdms.cwa.gov.tw/GeophyDownload.php}, 2012.
\newblock Data set.

\bibitem[Chang et~al.(2004)Chang, Cotton, Tsai, and Angelier]{Chang2004}
Tsui-Yu Chang, Fabrice Cotton, Yi-Ben Tsai, and Jacques Angelier.
\newblock Quantification of hanging-wall effects on ground motion: some insights from the 1999 chi-chi earthquake.
\newblock \emph{Bulletin of the Seismological Society of America}, 94\penalty0 (6):\penalty0 2186--2197, 2004.

\bibitem[Chen et~al.(2023)Chen, Yang, Wang, Li, Liu, and Beer]{Chen2023}
Guan Chen, Jiashu Yang, Ruohan Wang, Kaiqi Li, Yong Liu, and Michael Beer.
\newblock Seismic damage analysis due to near-fault multipulse ground motion.
\newblock \emph{Earthquake Engineering \& Structural Dynamics}, 52\penalty0 (15):\penalty0 5099--5116, 2023.

\bibitem[D’Amico et~al.(2019)D’Amico, Felicetta, Schiappapietra, Pacor, Gallovi{\v{c}}, Paolucci, Puglia, Lanzano, Sgobba, and Luzi]{DAmico2019}
Maria D’Amico, Chiara Felicetta, Erika Schiappapietra, Francesca Pacor, Franti{\v{s}}ek Gallovi{\v{c}}, Roberto Paolucci, Rodolfo Puglia, Giovanni Lanzano, Sara Sgobba, and Lucia Luzi.
\newblock Fling effects from near-source strong-motion records: Insights from the 2016 m w 6.5 norcia, central italy, earthquake.
\newblock \emph{Seismological Research Letters}, 90\penalty0 (2A):\penalty0 659--671, 2019.

\bibitem[Fenves and Ellery(1998)]{fenves1998behavior}
Gregory~L Fenves and Michael Ellery.
\newblock Behavior and failure analysis of a multiple-frame highway bridge in the 1994 northridge earthquake.
\newblock Technical report, PEER, 1998.

\bibitem[Graizer(2010)]{Graizer2010}
Vladimir Graizer.
\newblock Strong motion recordings and residual displacements: What are we actually recording in strong motion seismology?
\newblock \emph{Seismological Research Letters}, 81\penalty0 (4):\penalty0 635--639, 2010.

\bibitem[Graizer(1979)]{Graizer1979}
VM~Graizer.
\newblock Determination of the true displacement of the ground from strong-motion recordings. izv ussr acad sci.
\newblock \emph{Physics Solid Earth}, 15\penalty0 (12):\penalty0 875--885, 1979.

\bibitem[Iwan et~al.(1985)Iwan, Moser, and Peng]{Iwan1985SomeAccelerograph}
Wilfred~D. Iwan, Michael~A. Moser, and Chia-Yen Peng.
\newblock {Some observations on strong-motion earthquake measurement using a digital accelerograph}.
\newblock \emph{Bulletin of the Seismological Society of America}, 75\penalty0 (5):\penalty0 1225--1246, 10 1985.
\newblock ISSN 1943-3573.
\newblock \doi{10.1785/BSSA0750051225}.
\newblock URL \url{https://pubs.geoscienceworld.org/bssa/article/75/5/1225/102161/Some-observations-on-strong-motion-earthquake}.

\bibitem[Kazaz et~al.(2024)Kazaz, Bilge, and G{\"u}rb{\"u}z]{Kazaz2024}
{\.I}lker Kazaz, {\.I}smail~Hakk{\i} Bilge, and Muhammed G{\"u}rb{\"u}z.
\newblock Near-fault ground motion characteristics and its effects on a collapsed reinforced concrete structure in hatay during the february 6, 2023 mw7. 8 kahramanmara{\c{s}} earthquake.
\newblock \emph{Engineering Structures}, 298:\penalty0 117067, 2024.

\bibitem[Lay and Wallace(1995)]{Lay1995}
Thorne Lay and Terry~C Wallace.
\newblock \emph{Modern global seismology}, volume~58.
\newblock Elsevier, 1995.

\bibitem[Lee et~al.(2023)Lee, Liu, and Lin]{Lee2023}
Shiann-Jong Lee, Ting-Yu Liu, and Tzu-Chi Lin.
\newblock The role of the west-dipping collision boundary fault in the taiwan 2022 chihshang earthquake sequence.
\newblock \emph{Scientific Reports}, 13\penalty0 (1):\penalty0 3552, 2023.

\bibitem[Liao et~al.(2004)Liao, Loh, and Lee]{liao2004comparison}
Wen-I Liao, Chin-Hsiung Loh, and Bor-Han Lee.
\newblock Comparison of dynamic response of isolated and non-isolated continuous girder bridges subjected to near-fault ground motions.
\newblock \emph{Engineering Structures}, 26\penalty0 (14):\penalty0 2173--2183, 2004.

\bibitem[Loh et~al.(2002)Loh, Liao, and Chai]{Loh2002}
Chin-Hsiung Loh, Wen-I Liao, and Juin-Fu Chai.
\newblock Effect of near-fault earthquake on bridges: lessons learned from chi-chi earthquake.
\newblock \emph{Earthquake Engineering and Engineering Vibration}, 1:\penalty0 86--93, 2002.

\bibitem[Mavroeidis and Papageorgiou(2003)]{Mavroeidis2003}
George~P Mavroeidis and Apostolos~S Papageorgiou.
\newblock A mathematical representation of near-fault ground motions.
\newblock \emph{Bulletin of the seismological society of America}, 93\penalty0 (3):\penalty0 1099--1131, 2003.

\bibitem[Naeim et~al.(2000)Naeim, Lew, Huang, Lam, and Carpenter]{Naeim2000}
Farzad Naeim, Marshall Lew, Sampson~C Huang, Howard~K Lam, and Lauren~D Carpenter.
\newblock The performance of tall buildings during the 21 september 1999 chi-chi earthquake, taiwan.
\newblock \emph{The Structural Design of Tall Buildings}, 9\penalty0 (2):\penalty0 137--160, 2000.

\bibitem[Poulos and Miranda(2024)]{Poulos2024}
Alan Poulos and Eduardo Miranda.
\newblock Directionality characteristics of horizontal response spectra from the 2022 mw 6.9 chihshang, taiwan earthquake.
\newblock \emph{Earthquake Spectra}, 40\penalty0 (3):\penalty0 1683--1696, 2024.

\bibitem[Ramos-Sepulveda et~al.(2023)Ramos-Sepulveda, Parker, Thompson, Brandenberg, Li, Ilhan, Hashash, Rathje, and Stewart]{Ramos2023}
Mar{\'\i}a~E Ramos-Sepulveda, Grace~Alexandra Parker, Eric~M Thompson, Scott~J Brandenberg, Meibai Li, Okan Ilhan, Youssef~MA Hashash, Ellen~M Rathje, and Jonathan~P Stewart.
\newblock High-pass corner frequency selection for implementation in the usgs automated ground motion processing tool.
\newblock In \emph{Geo-Congress 2023}, pages 327--335, 2023.

\bibitem[Shabestari and Yamazaki(2003)]{Shabestari2003}
Khosrow~T Shabestari and Fumio Yamazaki.
\newblock Near-fault spatial variation in strong ground motion due to rupture directivity and hanging wall effects from the chi-chi, taiwan earthquake.
\newblock \emph{Earthquake engineering \& structural dynamics}, 32\penalty0 (14):\penalty0 2197--2219, 2003.

\bibitem[Shahi and Baker(2014)]{Shahi2014}
Shrey~K Shahi and Jack~W Baker.
\newblock An efficient algorithm to identify strong-velocity pulses in multicomponent ground motions.
\newblock \emph{Bulletin of the Seismological Society of America}, 104\penalty0 (5):\penalty0 2456--2466, 2014.

\bibitem[Shin and Teng(2001)]{Shin2001}
Tzay-Chyn Shin and Ta-liang Teng.
\newblock An overview of the 1999 chi-chi, taiwan, earthquake.
\newblock \emph{Bulletin of the seismological society of America}, 91\penalty0 (5):\penalty0 895--913, 2001.

\bibitem[Stewart et~al.(2024)Stewart, Buckreis, Shams, Mohammed, Meibai, Kishida, Lanzano, Luzi, Zimmaro, Bradley, Huang, Nweke, Brandenberg, Donahue, and Bozorgnia]{Stewart2024}
J.~P. Stewart, T.~E. Buckreis, R.~R. Shams, S.~Mohammed, L.~Meibai, T.~Kishida, G.~Lanzano, L.~Luzi, P.~Zimmaro, B.~A. Bradley, J.~Huang, C.~C. Nweke, S.~J. Brandenberg, J.~L. Donahue, and Y.~Bozorgnia.
\newblock Overview of the {NGA}\nobreakdash--west3 ground motion database.
\newblock In \emph{2024 SCEC Annual Meeting}, September 2024.
\newblock Poster presentation.

\bibitem[Tang et~al.(2023)Tang, Lin, Tung, Wang, Lee, Hsu, Shyu, Kuo, and Chen]{Tang2023}
Chi-Hsien Tang, Yunung~Nina Lin, Hsin Tung, Yu~Wang, Shiann-Jong Lee, Ya-Ju Hsu, J~Bruce~H Shyu, Yu-Ting Kuo, and Horng-Yue Chen.
\newblock Nearby fault interaction within the double-vergence suture in eastern taiwan during the 2022 chihshang earthquake sequence.
\newblock \emph{Communications Earth \& Environment}, 4\penalty0 (1):\penalty0 333, 2023.

\bibitem[Uzarski et~al.(1999)Uzarski, O'Rourke, Abrahamson, Amin, Goltz, Lam, and Tseng]{Uzarski1999ChiChi}
Joe Uzarski, Michael O'Rourke, Norman Abrahamson, Navin Amin, James Goltz, Ignatius~Po Lam, and Wen~S. Tseng.
\newblock The chi-chi, taiwan earthquake of september 21, 1999.
\newblock \emph{EERI Special Earthquake Report}, December 1999.
\newblock Learning from Earthquakes, Earthquake Engineering Research Institute.

\bibitem[Wang et~al.(2011)Wang, Schurr, Milkereit, Shao, and Jin]{Wang2011}
Rongjiang Wang, Bernd Schurr, Claus Milkereit, Zhigang Shao, and Mingpei Jin.
\newblock An improved automatic scheme for empirical baseline correction of digital strong-motion records.
\newblock \emph{Bulletin of the Seismological Society of America}, 101\penalty0 (5):\penalty0 2029--2044, 2011.

\bibitem[Yu and Gao(2001)]{Yu2001}
Yan-xiang Yu and Meng-tan Gao.
\newblock Effects of the hanging wall and footwall on peak acceleration during the jiji (chi-chi), taiwan province, earthquake.
\newblock \emph{Acta seismologica sinica}, 14\penalty0 (6):\penalty0 654--659, 2001.

\end{thebibliography}

\ifnum\flagreview>0
    \listoffigures
\fi

\end{document}